\documentclass[11pt]{article}
\usepackage[margin=0.9in]{geometry}
\usepackage{graphicx}
\usepackage[table]{xcolor}
\usepackage{amsmath,amssymb}
\usepackage{booktabs}
\usepackage{array}
\usepackage{tikz}
\usetikzlibrary{shapes.geometric,arrows.meta,positioning,fit,backgrounds}
\usepackage[colorlinks=true,linkcolor=moaBlue,citecolor=moaBlue,urlcolor=moaBlue]{hyperref}
\usepackage{tcolorbox}
\tcbuselibrary{skins,breakable}
\usepackage{fancyhdr}
\usepackage{titlesec}
\usepackage{authblk}

\definecolor{moaBlue}{HTML}{1B5FA8}
\definecolor{moaTeal}{HTML}{0E8C8C}
\definecolor{moaOrange}{HTML}{D2691E}
\definecolor{moaGreen}{HTML}{2E8B57}
\definecolor{moaPurple}{HTML}{6A3D9A}
\definecolor{moaGray}{HTML}{555555}
\definecolor{moaLightBlue}{HTML}{EAF2FB}
\definecolor{moaLightTeal}{HTML}{E6F7F5}
\definecolor{moaLightOrange}{HTML}{FBEFE2}
\definecolor{moaLightGreen}{HTML}{E9F5EC}
\definecolor{moaLightPurple}{HTML}{F1E9F8}

\titleformat{\section}{\Large\bfseries\color{moaBlue}}{\thesection}{0.6em}{}
\titleformat{\subsection}{\large\bfseries\color{moaTeal}}{\thesubsection}{0.6em}{}

\newtcolorbox{briefbox}[1][]{colback=moaLightBlue,colframe=moaBlue,fonttitle=\bfseries,title=In Brief,#1,breakable}
\newtcolorbox{findingbox}[1][]{colback=moaLightGreen,colframe=moaGreen,fonttitle=\bfseries,title=Real Finding,#1,breakable}
\newtcolorbox{opendbox}[1][]{colback=moaLightOrange,colframe=moaOrange,fonttitle=\bfseries,title=Open Question,#1,breakable}
\newtcolorbox{mechbox}[1][]{colback=moaLightPurple,colframe=moaPurple,fonttitle=\bfseries,title=Mechanism,#1,breakable}

\title{\color{moaBlue}\Large One Communicator, Three Vendors:\\
Extending the MoA Predictive Theory to a Real NVIDIA+AMD+CPU\\
MPI Ensemble}
\author[1]{Lenore M. Mullin\thanks{\texttt{lmullin@albany.edu}}}
\author[2]{Gaétan Hains\thanks{\texttt{gaetan.hains@u-pec.fr}}}
\affil[1]{College of Nanotechnology, Science and Engineering, University at Albany, SUNY, Albany, NY, USA}
\affil[2]{LACL, Université Paris-Est Créteil, Créteil, France}
\date{October 2026}

\begin{document}
\maketitle

\begin{abstract}
\noindent The companion MPI-extension paper validated the first real, communication-bearing
$\rho_{\mathrm{machine}}(d)$ ensemble: one NVIDIA A100 partition and one AMD-class CPU partition of a
single NCSA Delta node, under one genuine Open MPI job, reassembled by a real \texttt{MPI\_Gatherv}. That
result is still single-vendor on the accelerator side --- one GPU architecture, one NUMA-local CPU. This
paper merges that result with the heterogeneous multi-machine paper's multi-vendor, no-communication
partition (NVIDIA, AMD, and Intel devices, split across institutions with no shared interconnect) into one
theoretical and experimental design: a single Slurm \emph{heterogeneous job} spanning Delta's
\texttt{gpuA100x4} and \texttt{gpuMI100x8} partitions, placing NVIDIA A100, AMD MI100, and CPU ranks under
one real MPI communicator for the first time, with Intel's Data Center GPU Max 1550 (TACC Stampede3)
retained as the no-communication companion case its own institution's lack of shared interconnect still
requires. We give the theoretical extension in full --- the generalized $w_i \propto 1/t_i$ weight split
for three concurrent realizations, the admissibility of a Slurm het-job specification as a $C_d$-class
directive, and the hardware-placement mechanism itself --- report the first real results --- a
matching-shape AMD MI100 $t_d$; one working nine-rank communicator (4~A100, 4~CPU, 1~MI100) across both
het-job components, with its measured per-rank times and Gatherv cost; and an independent fp64 correctness
check --- and state plainly what the run exposed: the $w_i \propto 1/t_i$ split assumes time scales
linearly with rows, and that assumption fails for the MI100 rank (6~rows took 597~ms against a linear
prediction of 1.6~ms), so the slowest device sets the ensemble's time by a factor of about 90. We then give a launch-geometry
model and record its predictions before measuring them. The A100 test succeeds: its time is flat from 128 to
3456 query rows and steps at exactly 3520 and 6976 rows, as predicted. The MI100 shows a floor of about
600~ms; floor-decomposition runs place it in the key loop, and changing the kernel's launch settings (64 workers per team, keys split over 16 teams)
cut it to 14.6~ms and the energy per call at 6 rows by $32\times$, though the MI100 remains about $95\times$ slower than an A100 at 8192 rows. For the original
configuration the measured curves show that the CPU ranks (6.2~ms per-call floor) and the MI100 rank (about 600~ms) should receive no rows at this problem size, while the four A100
ranks finish in about 2.2~ms; the tuned MI100 would still receive none. we then ran that split as an ensemble and measured $2.21$~ms
(slowest compute plus Gatherv: $4.7$--$5.0$~ms in two runs and $9.1$~ms in a third, against $9.2$--$9.7$~ms for the $1/t$ split). A controlled set of reruns also shows that
the companion paper's A100 result ($8.1\times$ the linear prediction) and its $\approx22$~ms Gatherv were launch
artifacts: under \texttt{mpirun} with Open~MPI's default transport all four GPU ranks shared one A100
(Section~\ref{sec:onegpu}). Finally we derive from the shapes alone the message sizes
of a key/value-split (communication-bearing) partition, run it, and compare: the ring Allreduce of the
8.5~MB packed message took 11.3, 16.0 and 15.3~ms in three runs against a predicted 11--12~ms, the small Allreduce missed its
prediction in every run, and at this size the split is communication-bound. Repeating the jobs showed that compute times reproduce to about 1\% but communication times vary by up to a factor of four between runs. Each idea the new material depends on --- the core primitives, the DNF-to-ONF pipeline, and the
dimension-lifting step that turns a single-device derivation into a multi-device ensemble --- is given a
figure and a short, self-contained explanation first, and the new sections close with worked examples.
\end{abstract}

% ======================================================================
\section{MoA in Five Primitives}
\label{sec:primer}
% ======================================================================

MoA builds every array computation from a small, fixed set of primitive operators. Five recur constantly
throughout this paper series:

\begin{itemize}
\item $\psi$ (\textbf{psi}) --- indexing: picks out an element or sub-array by its index tuple.
\item $\iota$ (\textbf{iota}) --- index generation: produces the index set of an array from its shape.
\item $\rho$ (\textbf{rho}) --- shape: the sequence of an array's dimension sizes. The object that carries
every cost prediction in this paper, $\rho_{\mathrm{machine}}(d)$, is the \emph{shape of a machine}, in
direct analogy to the shape of an array.
\item $\mathrm{rav}$ (\textbf{ravel}) --- flattening: linearizes a multi-dimensional array into one
dimension, in a specified order. This is the operator that reassembles independently-computed device
outputs with no arithmetic combination, only concatenation --- the mechanism behind every
no-communication ensemble in this series.
\item $\gamma$ (\textbf{gamma}) --- a generalized indexing/reshaping operator used in tiling and blocking
derivations.
\end{itemize}

\begin{figure}[h]
\centering
\includegraphics[width=0.75\linewidth]{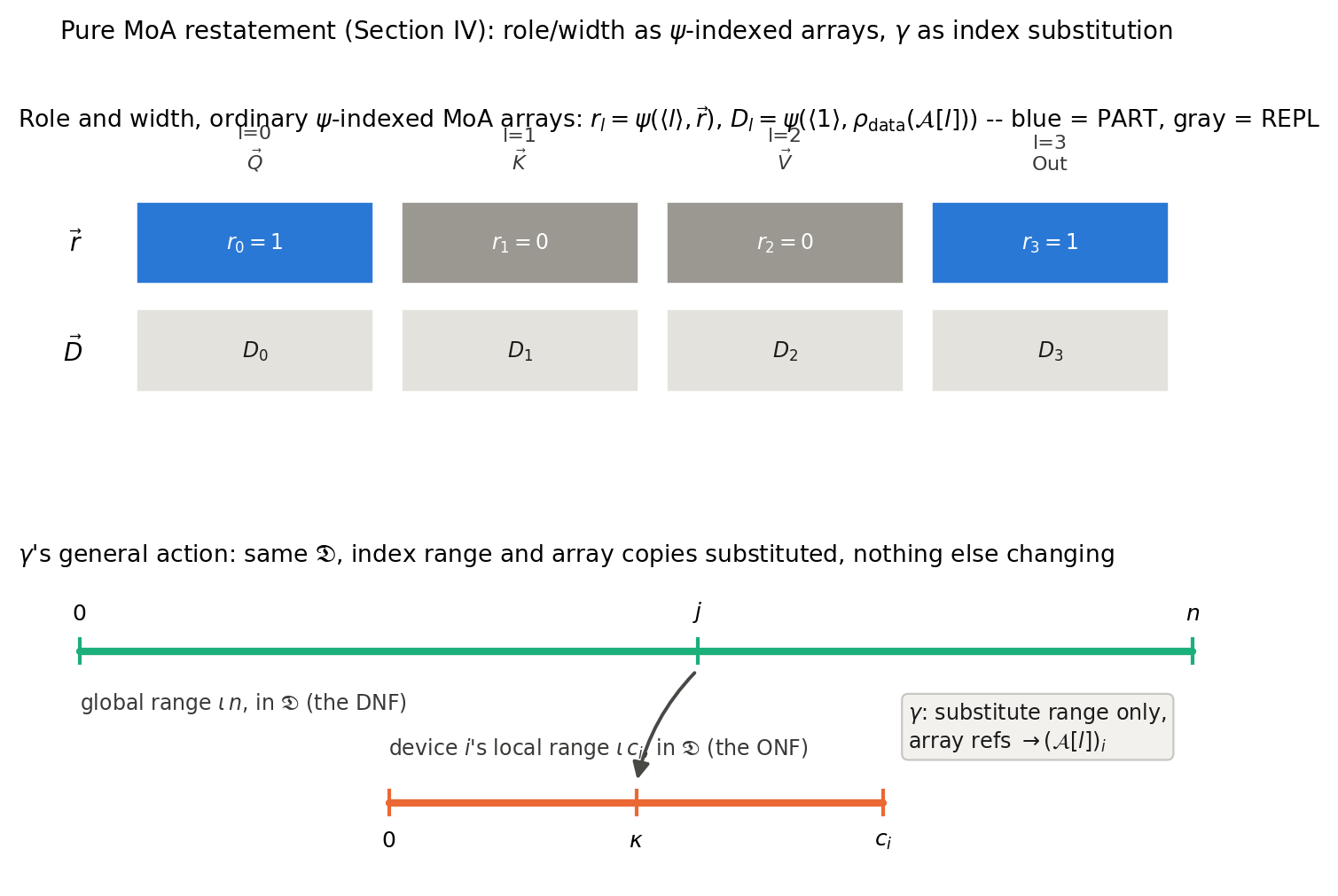}
\caption{The core MoA primitives and how they compose: shape ($\rho$) and index generation ($\iota$) feed
indexing ($\psi$); every tiling and dimension-lifting derivation in this paper builds on these same five
operators, whether the target is a single core, a single GPU, or a three-vendor MPI ensemble.}
\label{fig:pure-moa}
\end{figure}

\begin{briefbox}
A computation's \textbf{DNF} (Denotational Normal Form) says what the answer must be, with no mention of
tiles, threads, or memory hierarchy. Its \textbf{ONF} (Operational Normal Form) is where device-specific
decisions get made --- which tile size, how many threads, which precision. MoA's guarantee is that any
admissible ONF realizing a given DNF produces the same answer, provably, not just empirically. This paper
adds a \emph{third} vendor's ONF (AMD MI100, OpenMP target-offload) to a DNF this series has already
realized twice (NVIDIA A100 via Triton; a 15-core CPU via OpenMP) --- same DNF, same guarantee, a new
device-specific ONF.
\end{briefbox}

\begin{figure}[h]
\centering
\includegraphics[width=0.7\linewidth]{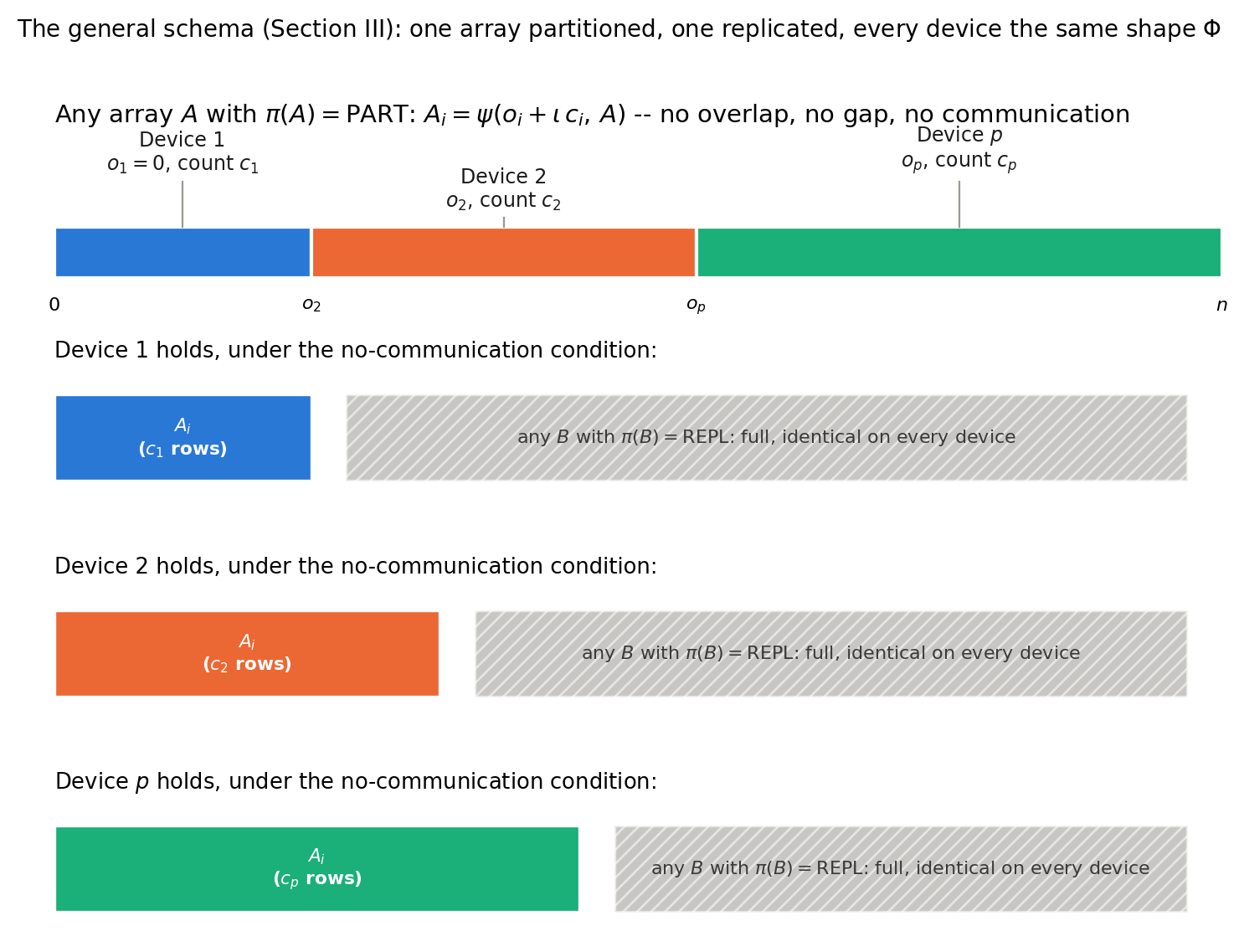}
\caption{One DNF specification, realized as a different ONF per target --- CPU, NVIDIA GPU (Triton), AMD
GPU (OpenMP target-offload), Intel GPU (OpenMP target-offload) --- each tuned to that device's own measured
shape, all provably computing the same answer. This paper is the first to put three of these four
realizations under one real MPI communicator at once.}
\label{fig:general-schema}
\end{figure}

\textbf{Dimension lifting} is the step that turns a single-device ONF into an ensemble: the query/batch
dimension of an attention computation is split across $d$ real machines, each with its own real, measured
shape $\rho_{\mathrm{machine}}(d)$, weighted by $w_i \propto 1/t_i$ --- faster devices get
proportionally more rows --- and the outputs are reassembled by $\mathrm{rav}$ (no communication) or by a
real collective such as \texttt{MPI\_Gatherv} (communication-bearing, the subject of this paper).

\begin{figure}[h]
\centering
\includegraphics[width=0.78\linewidth]{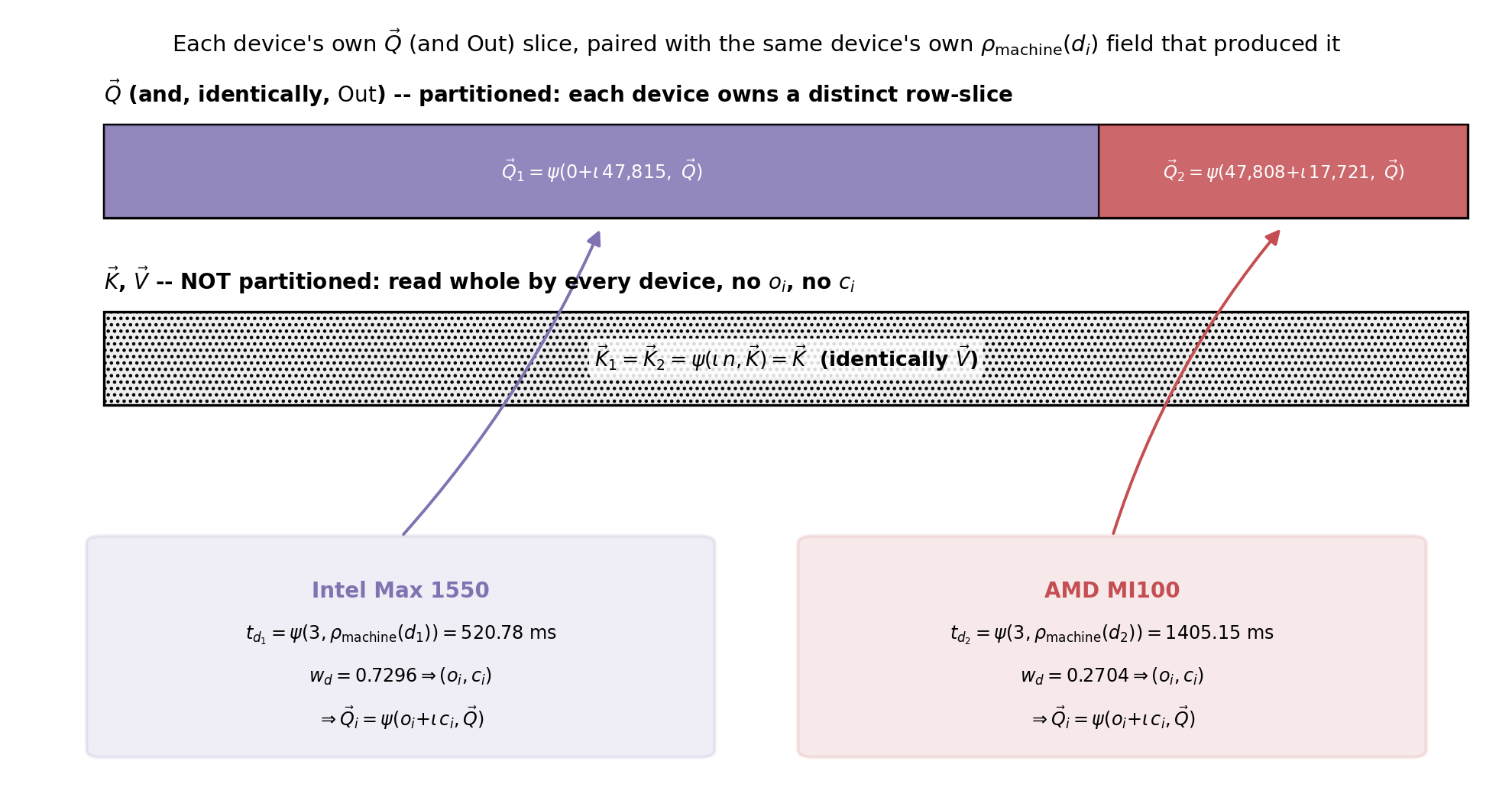}
\caption{Dimension lifting: an array's query dimension is partitioned by weight $w_i \propto 1/t_i$ across
$d$ real machines; each machine's own $\rho_{\mathrm{machine}}(d)$ determines its ONF and its share of the
work. The companion MPI-extension paper validated this for $d{=}2$ realizations (A100, CPU) under one real
communicator; this paper extends it to $d{=}3$.}
\label{fig:array-machine-mapping}
\end{figure}

% ======================================================================
\section{Where the Two Prior Lines Converge}
\label{sec:prior}
% ======================================================================

This paper sits at the intersection of two previously separate results in this series, and does not
restate either in full.

\begin{itemize}
\item The \textbf{heterogeneous multi-machine paper} \cite{hetero} validated a weighted, no-communication
partition across \emph{three vendors} (NVIDIA, AMD, Intel) and \emph{multiple institutions}: its first real
result split one attention problem between an Intel Max 1550 (TACC Stampede3) and an AMD MI100 (NCSA
Delta), two institutions with no shared interconnect, finishing within 7.5\% of each other by the
weighted-partition design. That result is multi-vendor but never shares a real MPI communicator --- each
device's job is submitted and timed independently, reassembly is by $\mathrm{rav}$ alone.
\item The \textbf{MPI-extension paper} \cite{mpiext} validated the opposite complement: a real, single
NCSA Delta node, every CPU and GPU under \emph{one} genuine Open MPI job, reassembled by a real
\texttt{MPI\_Gatherv}, with a measured $t_{\mathrm{comm}}$ and two new admissible fields ($G_d$, $L_d$).
That result is communication-bearing but single-vendor on the accelerator side --- 4$\times$ NVIDIA A100
and nothing else.
\end{itemize}

\textbf{This paper's contribution} is the natural next step neither prior paper took: a single real MPI
communicator spanning \emph{more than one accelerator vendor}. We bring AMD's MI100 into the same
communicator as the A100+CPU ensemble already validated, via a Slurm \emph{heterogeneous job} spanning
two different Delta partitions, and keep Intel's Max 1550 in the no-communication role \cite{hetero}
already gave it, since Stampede3 and Delta share no interconnect at all --- no mechanism this paper
proposes changes that physical fact.

% ======================================================================
\section{The Het-Job Mechanism: A Real, Admissible Directive}
\label{sec:hetjob}
% ======================================================================

A single \texttt{mpirun} job cannot ordinarily span two different Slurm partitions (\texttt{gpuA100x4} and
\texttt{gpuMI100x8} are different node types under one controller). Slurm's \textbf{heterogeneous job}
mechanism (\texttt{sbatch} with multiple \texttt{--hetjob} components, called via \texttt{\#SBATCH
hetjob} package directives or \texttt{sbatch : :} packjob syntax) is built for exactly this case: one job
allocation with two components, each requesting its own partition, node count, and GPU type, launched
together so Open MPI's PMI wire-up sees one combined set of ranks across both.

\begin{mechbox}
\textbf{Why this is admissible, not a heuristic.} Under \cite{closinggap}'s two-part criterion --- a
directive must select a real hardware quantity with no discretionary reordering, and be deterministic
rather than compiler/runtime-overridable --- a het-job component specification qualifies cleanly: each
component's \texttt{--partition}, \texttt{--nodes}, and \texttt{--gpus-per-node} are literal, honored
resource requests, not a scheduler heuristic the runtime is free to reinterpret. This extends $C_d$'s
admissible class (Section~\ref{sec:cd-extend}) to the job-orchestration layer, not just the
compiler/runtime-directive layer the prior two papers covered.
\end{mechbox}

\begin{findingbox}
\textbf{How the ranks can actually be launched across the components on Delta.} Several runs of the
het-job script and four short probe jobs established the following on the real system.
\begin{itemize}
\item \emph{Allocation works.} Slurm accepts the two-component job and starts both components together
(an A100 node and \texttt{gpud01}); \texttt{SLURM\_JOB\_NODELIST\_HET\_GROUP\_0/1} resolve to the two
hosts, and the MI100 host is reachable by \texttt{ping} from the A100 node.
\item \emph{A single \texttt{mpirun} with a combined rankfile cannot span the components.}
\texttt{SLURM\_JOB\_NODELIST} lists only component~0, so Open MPI rejected the MI100 host as ``not
allocated'' (exit status 213) before any rank started.
\item \emph{\texttt{srun} multiple-program launch can.} A single \texttt{srun --mpi=pmix\_v5 -n 8 \ldots{} : -n 1
\ldots{}} step places global ranks 0--7 on component~0 and rank~8 on component~1, the layout the weighted
driver expects. Per-rank core pinning, previously done by the rankfile, moves into a small per-task
wrapper (\texttt{taskset}).
\item \emph{PMIx client mismatch, resolved.} The first \texttt{srun} launch failed inside
\texttt{MPI\_Init} with \texttt{PMIX\_ERR\_FILE\_OPEN\_FAILURE} (shared-memory data store of the PMIx library
bundled with the HPC-X Open MPI module). Setting \texttt{PMIX\_MCA\_gds=hash} and
\texttt{PMIX\_MCA\_psec=native} removed that failure.
\item \emph{Cancellations tied to the UCX transport.} Four \texttt{srun --mpi=pmix\_v5} launches that let
Open MPI select UCX were terminated as a whole job, roughly five seconds after the ranks started and before
any rank printed (\texttt{sacct}: ``CANCELLED by~0''): two full het-job attempts, a nine-rank probe across
both components, and an 8-rank step confined to a single A100 node, so the cause is not the span across
components. (An \texttt{mpirun} on one node with UCX available, and an \texttt{srun --mpi=pmi2} launch that
started independent single-rank processes, were not cancelled.) Every launch with UCX excluded
(\texttt{OMPI\_MCA\_pml=ob1}, \texttt{OMPI\_MCA\_btl=self,vader,tcp}, and the UCC collective component,
which sits on UCX, disabled) ran to completion, on one node and across both components, under
\texttt{srun} and under \texttt{mpirun} over \texttt{ssh}. We have not identified why a UCX start-up leads to a
root-issued cancellation; an NCSA ticket is open. The effect is that inter-partition traffic here runs
over TCP.
\end{itemize}
\textbf{Status.} Confirmed: one communicator of nine ranks spans both components, with the MI100 rank on
\texttt{gpud01}. The results are in the next section.
\end{findingbox}

\begin{center}
\resizebox{0.97\linewidth}{!}{%
\begin{tikzpicture}[
  node distance=6mm and 8mm,
  every node/.style={font=\small},
  nvbox/.style={draw=moaBlue, thick, fill=moaLightBlue, rounded corners, minimum width=2.6cm, minimum height=0.9cm, align=center},
  amdbox/.style={draw=moaOrange, thick, fill=moaLightOrange, rounded corners, minimum width=2.6cm, minimum height=0.9cm, align=center},
  cpubox/.style={draw=moaTeal, thick, fill=moaLightTeal, rounded corners, minimum width=2.6cm, minimum height=0.9cm, align=center},
  intelbox/.style={draw=moaPurple, thick, dashed, fill=moaLightPurple, rounded corners, minimum width=2.6cm, minimum height=0.9cm, align=center},
  hetbox/.style={draw=moaGray, dashed, rounded corners, inner sep=10pt},
  lbl/.style={font=\scriptsize\color{moaGray}}
]
\node[nvbox] (g0) {A100 $\times$4\\[1pt]\scriptsize NVIDIA, ranks 0-3};
\node[cpubox, below=of g0] (c0) {CPU $\times$4\\[1pt]\scriptsize ranks 4-7};
\node[amdbox, right=62mm of g0] (m0) {MI100 $\times$k\\[1pt]\scriptsize AMD, ranks 8-(7+k)};

\node[hetbox, fit=(g0)(c0), label={[lbl]above left:het-job component 0: gpuA100x4}] (het0) {};
\node[hetbox, fit=(m0), label={[lbl]above left:het-job component 1: gpuMI100x8}] (het1) {};

\node[intelbox, right=40mm of m0] (intel) {Max 1550\\[1pt]\scriptsize Intel, Stampede3};

\draw[-{Latex[length=2mm]}, moaGray, thick] (het0.east) -- (het1.west) node[midway, above=1mm, lbl, align=center] {one real\\MPI\_COMM\_WORLD};
\draw[-{Latex[length=2mm]}, moaGray, dashed, thick] (het1.east) -- (intel.west) node[midway, above=2mm, lbl] {no shared interconnect};
\end{tikzpicture}%
}
\end{center}
\begin{center}
\small\color{moaGray} The design this paper validates theoretically and proposes experimentally: one real
Open MPI communicator spans a Slurm het-job's two components (NVIDIA A100 + CPU, AMD MI100), solid line;
Intel's Max 1550 remains the no-communication companion case, dashed line, since Stampede3 shares no
interconnect with Delta.
\end{center}

% ======================================================================
\section{Extending the Weight Formula to Three Realizations}
\label{sec:weights}
% ======================================================================

The companion MPI-extension paper's weight split generalizes directly. For $d$ concurrent realizations,
each with its own real, measured full-problem time $t_i$ at matching shape:
\[
w_i = \frac{1/t_i}{\sum_{j=1}^{d} 1/t_j}, \qquad \text{rows}_i = \lfloor w_i \cdot N \rfloor,
\]
with any rounding remainder absorbed by a fixed, deterministic tie-break (the companion paper's own
convention: the fastest rank absorbs it) so every rank computes the identical split with no communication.
This formula is already validated for $d{=}2$ realization types (A100, CPU) \cite{mpiext}; nothing in its
derivation assumes $d{=}2$ specifically. What is new here is not the formula but a fourth necessary input:
a real, measured AMD MI100 $t_d$ at the \emph{same} matching shape ($N{=}8192$, $D{=}64$, $B{\cdot}H{=}2$)
the A100 ($6.874$~ms) and CPU ($221.031$~ms) numbers were measured at \cite{mpiext}.

\begin{findingbox}
\textbf{A real MI100 $t_d$ at the matching shape.} The standalone test
(\texttt{delta\_mi100\_matching\_shape\_standalone\_test.slurm}) ran $N{=}8192$, $D{=}64$, $B{\cdot}H{=}2$,
fp32, full length (\texttt{QUERY\_OFFSET=0}, \texttt{QUERY\_CHUNK=}$N$) on a node verified as \texttt{gfx908}
before compilation (the script aborts otherwise, so a host-only fallback cannot masquerade as a GPU time),
with the real derived $C_d$ pins \texttt{NUM\_WORKERS=10}, \texttt{VECTOR\_LENGTH=64}, \texttt{BLOCK\_N=16}
\cite{closinggap}. Two independent runs gave $2136.100$~ms and $2132.421$~ms (a $0.17\%$ spread); we use
their mean, $t_{\mathrm{MI100}}=2134.261$~ms. Both runs matched the CPU reference
($\max|\mathrm{err}|=5.1260\times10^{-6}$), and \texttt{rocm-smi} showed device utilization rising from
$0\%$ to $100\%$ through the timed loop (package power from about 38~W idle to as high as about 170~W), confirming execution on
the MI100. The device energy counter reported a zero delta under that confirmed load, the same behavior
\cite{closinggap} documented for this device; no MI100 energy figure is claimed here.
\end{findingbox}

\begin{table}[h]
\centering
\renewcommand{\arraystretch}{1.3}
\small
\begin{tabular}{@{}p{2.9cm}p{3.7cm}p{1.2cm}p{1.8cm}p{1.9cm}p{2.1cm}@{}}
\toprule
\textbf{Realization} & \textbf{Implementation} & \textbf{Prec.} & \textbf{$t_d$ (ms)} &
\textbf{vs.\ A100} & \textbf{Rows/rank} \\
\midrule
\rowcolor{moaLightBlue} NVIDIA A100 & Triton & fp32 & 6.874 & $1\times$ & 1984 (rank~0: 1990) \\
\rowcolor{moaLightTeal} CPU, 15 cores & OpenMP & fp64 & 221.031 & $32\times$ & 61 \\
\rowcolor{moaLightOrange} AMD MI100 & OpenMP target-offload & fp32 & 2134.261 & $310\times$ & 6 \\
\bottomrule
\end{tabular}
\caption{Matching-shape full-problem times and the resulting $w_i \propto 1/t_i$ row split for 4~A100, 4~CPU
and 1~MI100 rank ($4\cdot1984+4\cdot61+6=8186$ rows, the remainder of 6 absorbed by rank~0).}
\label{tab:three-t}
\end{table}

At the pinned $C_d$, the MI100 realization is about $9.7\times$ slower than the 15-core CPU realization, so
the weight formula assigns the single MI100 rank only 6 of 8192 rows ($0.07\%$). The weights thus ask
the MI100 rank to do very little; the measured run in Section~\ref{sec:phase3} shows that even that little
is too much. This is the time of
the realization \emph{as currently pinned}, not a claim about the device's best reachable time: the
closing-the-gap paper \cite{closinggap} found that the occupancy cap has to be re-derived for each
admissible width realization, and the pins used here were carried over from that paper's experiment rather
than re-derived for this shape. The kernel already carries the compile-time
\texttt{QUERY\_OFFSET}/\texttt{QUERY\_CHUNK} row-slicing interface the CPU kernel used in
\cite{mpiext}, so each MI100 rank owns a distinct row range the same deterministic way as every other rank
in this series.

% ======================================================================
\section{The First Three-Vendor Run}
\label{sec:phase3}
% ======================================================================

\begin{findingbox}
\textbf{One communicator, three vendors, and a linear-scaling assumption that fails.}
Nine ranks (4~A100 via Triton fp32, 4~CPU NUMA-domain ranks via OpenMP fp64, 1~MI100 via OpenMP
target-offload fp32) formed one MPI communicator across Delta's \texttt{gpuA100x4} and
\texttt{gpuMI100x8} partitions and computed the $N{=}8192$, $D{=}64$, $B{\cdot}H{=}2$ attention problem with
the row split of Table~\ref{tab:three-t}, launched by one \texttt{srun} multiple-program step with UCX
excluded (Section~\ref{sec:hetjob}). The stitched output has the expected shape $(2,8192,64)$, no NaNs and no
empty rows, and matches an independent fp64 reference to $\max|\mathrm{err}|=5.2\times10^{-7}$ over all
nine row blocks.
\end{findingbox}

\begin{table}[h]
\centering
\renewcommand{\arraystretch}{1.3}
\small
\begin{tabular}{@{}p{2.2cm}p{2.0cm}p{2.3cm}p{2.6cm}p{1.6cm}p{2.4cm}@{}}
\toprule
\textbf{Rank type} & \textbf{Rows/rank} & \textbf{Measured (ms)} & \textbf{Linear pred.\ (ms)} &
\textbf{Ratio} & \textbf{$\max|\mathrm{err}|$} \\
\midrule
\rowcolor{moaLightBlue} A100 $\times4$ & 1984 (1990) & 2.21--2.33 & 1.67 & $1.3$--$1.4\times$ & $5.2\times10^{-7}$ \\
\rowcolor{moaLightTeal} CPU $\times4$ & 61 & 6.63--6.83 & 1.65 & $4.0$--$4.2\times$ & $6.5\times10^{-16}$ \\
\rowcolor{moaLightOrange} MI100 $\times1$ & 6 & 597.0 & 1.56 & $382\times$ & $3.3\times10^{-7}$ \\
\bottomrule
\end{tabular}
\caption{Measured per-rank compute (median of 5 timed repeats after 3 warm-ups, all nine ranks running
concurrently) against the linear-scaling prediction $(\mathrm{rows}_i/N)\,t_i$. One job, one set of
measurements; no run-to-run spread is available yet.}
\label{tab:phase3}
\end{table}

\begin{center}
\resizebox{0.97\linewidth}{!}{%
\begin{tikzpicture}[x=1cm,y=0.62cm]
\node[anchor=east,font=\scriptsize] at (-0.15,0) {rank 0 A100, 1990 rows};
\fill[moaBlue!35,draw=moaBlue,thick] (0,-0.32) rectangle (1.099,0.32);
\draw[moaGray,very thick] (0.668,-0.42) -- (0.668,0.42);
\node[anchor=west,font=\scriptsize] at (1.219,0) {2.3~ms};
\node[anchor=east,font=\scriptsize] at (-0.15,-1) {rank 1 A100, 1984 rows};
\fill[moaBlue!35,draw=moaBlue,thick] (0,-1.32) rectangle (1.034,-0.6799999999999999);
\draw[moaGray,very thick] (0.664,-1.42) -- (0.664,-0.5800000000000001);
\node[anchor=west,font=\scriptsize] at (1.154,-1) {2.2~ms};
\node[anchor=east,font=\scriptsize] at (-0.15,-2) {rank 2 A100, 1984 rows};
\fill[moaBlue!35,draw=moaBlue,thick] (0,-2.32) rectangle (1.031,-1.68);
\draw[moaGray,very thick] (0.664,-2.42) -- (0.664,-1.58);
\node[anchor=west,font=\scriptsize] at (1.151,-2) {2.2~ms};
\node[anchor=east,font=\scriptsize] at (-0.15,-3) {rank 3 A100, 1984 rows};
\fill[moaBlue!35,draw=moaBlue,thick] (0,-3.32) rectangle (1.044,-2.68);
\draw[moaGray,very thick] (0.664,-3.42) -- (0.664,-2.58);
\node[anchor=west,font=\scriptsize] at (1.164,-3) {2.2~ms};
\node[anchor=east,font=\scriptsize] at (-0.15,-4) {rank 4 CPU, 61 rows};
\fill[moaTeal!35,draw=moaTeal,thick] (0,-4.32) rectangle (2.477,-3.68);
\draw[moaGray,very thick] (0.649,-4.42) -- (0.649,-3.58);
\node[anchor=west,font=\scriptsize] at (2.597,-4) {6.7~ms};
\node[anchor=east,font=\scriptsize] at (-0.15,-5) {rank 5 CPU, 61 rows};
\fill[moaTeal!35,draw=moaTeal,thick] (0,-5.32) rectangle (2.465,-4.68);
\draw[moaGray,very thick] (0.649,-5.42) -- (0.649,-4.58);
\node[anchor=west,font=\scriptsize] at (2.585,-5) {6.6~ms};
\node[anchor=east,font=\scriptsize] at (-0.15,-6) {rank 6 CPU, 61 rows};
\fill[moaTeal!35,draw=moaTeal,thick] (0,-6.32) rectangle (2.476,-5.68);
\draw[moaGray,very thick] (0.649,-6.42) -- (0.649,-5.58);
\node[anchor=west,font=\scriptsize] at (2.596,-6) {6.7~ms};
\node[anchor=east,font=\scriptsize] at (-0.15,-7) {rank 7 CPU, 61 rows};
\fill[moaTeal!35,draw=moaTeal,thick] (0,-7.32) rectangle (2.503,-6.68);
\draw[moaGray,very thick] (0.649,-7.42) -- (0.649,-6.58);
\node[anchor=west,font=\scriptsize] at (2.623,-7) {6.8~ms};
\node[anchor=east,font=\scriptsize] at (-0.15,-8) {rank 8 MI100, 6 rows};
\fill[moaOrange!35,draw=moaOrange,thick] (0,-8.32) rectangle (8.328,-7.68);
\draw[moaGray,very thick] (0.582,-8.42) -- (0.582,-7.58);
\node[anchor=west,font=\scriptsize] at (8.448,-8) {597~ms};
\draw[->,moaGray] (0,-9.0) -- (9.4,-9.0);
\foreach \x/\l in {0/1,3/10,6/100,9/1000}{\draw[moaGray] (\x,-8.9) -- (\x,-9.1) node[below,font=\scriptsize]{\l};}
\node[font=\scriptsize,anchor=west] at (9.5,-9.0) {ms (log scale)};
\draw[moaGray,very thick] (6.2,-10.15) -- (6.2,-10.75);
\node[font=\scriptsize,anchor=west] at (6.35,-10.45) {linear-scaling prediction};
\end{tikzpicture}%
}
\end{center}
\begin{center}
\small\color{moaGray} Measured compute time per rank, log scale (bars), against each rank's linear-scaling
prediction (dark tick). The eight A100 and CPU ranks sit within a factor of about 4 of their prediction; the
one MI100 rank, given 6 rows, sits a factor of 382 above it.
\end{center}

\textbf{What the numbers say.} The weight formula treats $t_i$ as a per-row cost: a device given
$\mathrm{rows}_i = w_i N$ rows should finish in about $w_i t_i$. That holds to within a small factor for
the A100 and CPU ranks. It does not hold for the MI100: its full-problem time of 2134~ms was measured with
all 8192 query rows in flight, and a 6-row slice still takes 597~ms, so the cost behaves like a large
fixed term plus a small per-row term, not like a rate. The CPU ranks show a smaller fixed term (about 6.2~ms).
The companion paper \cite{mpiext} reported the same breakdown for its A100 ranks ($8.1\times$, $13.5$~ms) and
attributed it to per-invocation overhead. We withdraw that attribution for the A100: the figure measured here
($1.3$--$1.4\times$, $2.2$~ms, passing fp64 check) reproduces in a separate scan and in an A100-only ensemble
(2.21~ms), and Section~\ref{sec:onegpu} shows that the $13.5$~ms came from four ranks sharing one GPU. The
CPU floor remains real and open; the MI100 floor is a property of its kernel's launch settings (Section~\ref{sec:tiles}). Taking the slowest rank's compute plus the measured
Gatherv gives an ensemble iteration of about $597+4.0\approx601$~ms; had every rank followed its linear
prediction, the same expression gives about $1.7+4.0\approx5.7$~ms. Without the MI100 rank the slowest
compute is the CPU ranks' $6.8$~ms. We do not report the driver's wall-clock total ($6.80$~s) as an ensemble
time: it is dominated by the A100 ranks' one-time Triton compilation during warm-up. Why a 6-row slice is
so slow on the MI100 is not established; Section~\ref{sec:tiles} tests a launch-geometry explanation read from
the kernels against row scans on every device.

\textbf{Communication.} The Gatherv, timed after a barrier so it excludes waiting for slower ranks, took
at most $4.03$~ms across the nine ranks, with Open MPI's \texttt{ob1} point-to-point layer over TCP between
the two partitions (UCX excluded). A separate 8~MiB ping-pong between rank~0 and the MI100 rank in a probe
job gave a round trip of about $8.4$~ms, roughly $2$~GB/s one way; this is a single rough measurement; Section~\ref{sec:comm} measures it properly. We
do not yet compare this $t_{\mathrm{comm}}$ with the two-vendor $\approx$22~ms of \cite{mpiext}: the
transport differs (TCP here), and the timing definition must first be confirmed identical.

\begin{mechbox}
\textbf{MoA in one worked example: how nine outputs become one array.} Each rank computes a block of rows,
which is $\psi$ (indexing) applied to the output with the index set $\iota$ of its rows. For the MI100
rank, $\iota(6)$ offset by $8186$ selects rows $8186,\dots,8191$. Because every block is a contiguous run of
the row-major $\mathrm{rav}$ (ravel) of the output, joining the blocks in rank order \emph{is} the ravel of
the full array --- no arithmetic, only placement. The Gatherv does exactly this: rank~8 contributes
$6\cdot2\cdot64=768$ values starting at element $8186\cdot2\cdot64=1{,}047{,}808$. This is the same
no-communication reassembly $\mathrm{rav}$ provides in every earlier paper of this series, now carried by a
real collective across three vendors.
\end{mechbox}

\begin{center}
\resizebox{0.97\linewidth}{!}{%
\begin{tikzpicture}[x=1cm,y=1cm,font=\small]
% full-length strip, to scale: 8192 rows over 15 cm
\def\sc{0.0018311}
\fill[moaBlue!35,draw=moaBlue,thick] (0,0) rectangle (1990*\sc,0.7);
\fill[moaBlue!25,draw=moaBlue,thick] (1990*\sc,0) rectangle (3974*\sc,0.7);
\fill[moaBlue!35,draw=moaBlue,thick] (3974*\sc,0) rectangle (5958*\sc,0.7);
\fill[moaBlue!25,draw=moaBlue,thick] (5958*\sc,0) rectangle (7942*\sc,0.7);
\fill[moaTeal!40,draw=moaTeal,thick] (7942*\sc,0) rectangle (8186*\sc,0.7);
\fill[moaOrange!70,draw=moaOrange,thick] (8186*\sc,0) rectangle (8192*\sc,0.7);
\foreach \a/\b/\t in {0/1990/{rank 0: A100},1990/3974/{rank 1},3974/5958/{rank 2},5958/7942/{rank 3}}{
  \node[font=\scriptsize] at ({(\a+\b)/2*\sc},0.35) {\t};}
\node[font=\scriptsize,anchor=south] at (7.5,0.8) {\textbf{rav of the output, rows $0$--$8191$ (to scale)}};
% zoom
\draw[moaGray,dashed] (7942*\sc,0) -- (11.0,-1.4);
\draw[moaGray,dashed] (8192*\sc,0) -- (15.0,-1.4);
\foreach \i/\c/\l in {0/moaTeal/{r4: 61},1/moaTeal/{r5: 61},2/moaTeal/{r6: 61},3/moaTeal/{r7: 61},4/moaOrange/{r8: 6}}{
  \fill[\c!40,draw=\c,thick] ({11.0+\i*0.8},-2.1) rectangle ({11.0+\i*0.8+0.8},-1.4);
  \node[font=\tiny] at ({11.0+\i*0.8+0.4},-1.75) {\l};}
\node[font=\scriptsize,anchor=north] at (13.0,-2.15) {the last 250 rows, enlarged (not to scale)};
\end{tikzpicture}%
}
\end{center}
\begin{center}
\small\color{moaGray} The output array as the Gatherv assembles it. Blue: the four A100 ranks (about 24\% of
the rows each); teal: the four CPU ranks (61 rows each); orange: the single MI100 rank (6 rows, $0.07\%$ of
the array, too thin to see at scale). The assembly is $\mathrm{rav}$ concatenation in rank order.
\end{center}

% ======================================================================
\section{The Shape of a Machine and the Shape of the Work}
\label{sec:shapes}
% ======================================================================

MoA's first move is to ask for the \emph{shape} of an array: the sequence of its dimension sizes, $\rho A$. The
attention output here has shape $\rho\,\mathrm{Out}=\langle 2,\,8192,\,64\rangle$: two problems ($B\!\cdot\!H$), 8192
query rows, 64 columns. This paper's claim is that a machine has a shape too, written $\rho_{\mathrm{machine}}(d)$, and that
cost is decided by how the shape of the work meets the shape of the machine. We have used that phrase throughout; here are
the shapes themselves, as read from the hardware, the launch scripts and the kernels (Figure~\ref{fig:machine-shapes}).

\begin{table}[h]
\centering
\renewcommand{\arraystretch}{1.3}
\small
\begin{tabular}{@{}p{2.2cm}p{4.4cm}p{3.9cm}p{2.9cm}@{}}
\toprule
\textbf{Machine} & \textbf{Shape $\rho_{\mathrm{machine}}$ (outermost first)} & \textbf{Unit the kernel launches} & \textbf{Source} \\
\midrule
\rowcolor{moaLightBlue} NVIDIA A100 (one of 4 per node) & $\langle 108\rangle$ streaming multiprocessors &
Triton program: a $64$-row $\times$ $64$-key tile; grid shape $\langle 2,\lceil r/64\rceil\rangle$ & kernel file; $108$ SMs\\
\rowcolor{moaLightTeal} CPU, one node & $\langle 4,\,16\rangle$: four NUMA domains of 16 cores; each domain is 1 GPU-rank core + a
15-core OpenMP team & OpenMP thread team of 15; rows divided among threads & rank--core map in the run logs\\
\rowcolor{moaLightOrange} AMD MI100 (one used) & $\langle 120\rangle$ compute units &
OpenMP team of 10 workers: a $32$-row tile; grid shape $\langle 2,\lceil r/32\rceil\rangle$, each team walking all 512 key tiles & kernel source; $120$ CUs (vendor figure)\\
\bottomrule
\end{tabular}
\caption{Machine shapes. The grid shape of a GPU launch is itself a shape: $\langle B\!\cdot\!H,\lceil r/\mathrm{tile}\rceil\rangle$, and
the number of programs (teams) is the product of its entries.}
\label{tab:machineshapes}
\end{table}

\begin{figure}[h]
\centering
\resizebox{0.98\linewidth}{!}{%
\begin{tikzpicture}[x=1cm,y=1cm,font=\small]
% A100: 12 x 9 = 108 SMs, first 62 filled (1984 rows)
\node[font=\small\bfseries] at (2.3,5.7) {A100: $\langle108\rangle$ SMs};
\foreach \i in {0,...,107}{
  \pgfmathtruncatemacro{\cx}{mod(\i,12)}
  \pgfmathtruncatemacro{\cy}{int(\i/12)}
  \ifnum\i<62 \fill[moaBlue!55,draw=white] ({\cx*0.4},{4.9-\cy*0.4}) rectangle ++(0.36,0.36);
  \else \fill[moaLightBlue,draw=white] ({\cx*0.4},{4.9-\cy*0.4}) rectangle ++(0.36,0.36);\fi}
\node[font=\scriptsize,align=center] at (2.3,0.75) {1984 rows $\Rightarrow$ 62 programs\\ 62 of 108 SMs busy: one wave};
% CPU: 4 NUMA boxes of 16 cores
\node[font=\small\bfseries] at (8.1,5.7) {CPU: $\langle4,16\rangle$};
\foreach \n in {0,1,2,3}{
  \pgfmathsetmacro{\ox}{6.2+mod(\n,2)*2.0}
  \pgfmathsetmacro{\oy}{4.9-int(\n/2)*2.2}
  \draw[moaTeal,thick] (\ox-0.05,\oy-1.45) rectangle (\ox+1.75,\oy+0.5);
  \foreach \c in {0,...,15}{
    \pgfmathtruncatemacro{\cx}{mod(\c,4)}
    \pgfmathtruncatemacro{\cy}{int(\c/4)}
    \ifnum\c=0 \fill[moaGray!60,draw=white] ({\ox+\cx*0.42},{\oy+0.1-\cy*0.4}) rectangle ++(0.38,0.36);
    \else \fill[moaTeal!55,draw=white] ({\ox+\cx*0.42},{\oy+0.1-\cy*0.4}) rectangle ++(0.38,0.36);\fi}}
\node[font=\scriptsize,align=center] at (8.1,0.75) {grey: GPU-rank core\\ teal: 15-core OpenMP team\\ 61 rows per team};
% MI100: 12 x 10 = 120 CUs, 2 filled (6 rows)
\node[font=\small\bfseries] at (13.6,5.7) {MI100: $\langle120\rangle$ CUs};
\foreach \i in {0,...,119}{
  \pgfmathtruncatemacro{\cx}{mod(\i,12)}
  \pgfmathtruncatemacro{\cy}{int(\i/12)}
  \ifnum\i<2 \fill[moaOrange!70,draw=white] ({11.3+\cx*0.4},{5.2-\cy*0.4}) rectangle ++(0.36,0.36);
  \else \fill[moaLightOrange,draw=white] ({11.3+\cx*0.4},{5.2-\cy*0.4}) rectangle ++(0.36,0.36);\fi}
\node[font=\scriptsize,align=center] at (13.6,0.75) {6 rows $\Rightarrow$ 2 teams\\ 2 of 120 CUs busy,\\ yet about 600 ms};
\end{tikzpicture}%
}
\end{figure}
\begin{center}
\small\color{moaGray}\refstepcounter{figure}\label{fig:machine-shapes}Figure~\thefigure. The shape of each machine, drawn as its units, with the units the
work occupies filled. Occupancy is not what makes the MI100 slow: two of its 120 compute units are busy and the call still takes
about 600~ms, so the floor comes from something other than how many units the work fills.
\end{center}

\begin{briefbox}[title={Worked example: where the A100 steps come from, from the shapes alone}]
The work has shape $\langle2,\,r,\,64\rangle$. The A100 kernel cuts the $r$ rows into tiles of 64, so the grid of tiles has
shape $\langle 2,\lceil r/64\rceil\rangle$, and the index set of that grid is
$\iota\langle 2,\lceil r/64\rceil\rangle$. The number of programs is the product of the entries of the grid shape. The machine
runs at most 108 at a time (one per SM), so the number of waves is $\lceil \text{programs}/108\rceil$.
\begin{center}
\small
\begin{tabular}{@{}rcccc@{}}
\toprule
$r$ & grid shape & programs & waves & measured (ms) \\
\midrule
\rowcolor{moaLightBlue} 1984 & $\langle2,31\rangle$ & 62 & 1 & 2.21 \\
\rowcolor{moaLightBlue} 3456 & $\langle2,54\rangle$ & 108 & 1 & 2.17 \\
\rowcolor{moaLightOrange} 3520 & $\langle2,55\rangle$ & 110 & 2 & 4.03 \\
\rowcolor{moaLightOrange} 6912 & $\langle2,108\rangle$ & 216 & 2 & 4.27 \\
\rowcolor{moaLightGreen} 6976 & $\langle2,109\rangle$ & 218 & 3 & 6.02 \\
\bottomrule
\end{tabular}
\end{center}
One more row of 64 pushes 3456 rows ($108$ programs, exactly the machine's width) to 3520 rows ($110$ programs), and the time
doubles; the same happens between 6912 and 6976. The two steps were predicted before the scan from nothing but these two
shapes (the grid and $\langle108\rangle$). The measured heights are about 12\% below the prediction, which is the part the
shapes alone do not give us.
\end{briefbox}

The same reading applies to the MI100 grid $\langle 2,\lceil r/32\rceil\rangle$, with one difference that the shapes also
predict: each of its teams walks all 512 key tiles in order, so the work per team does not shrink with the number of rows.
That is why the MI100's time cannot be written as rows times a rate, and why the floor experiment below looks inside the
loop.

% ======================================================================
\section{A Correction: Four Ranks on One GPU}
\label{sec:onegpu}
% ======================================================================

The companion paper's final run (job 22603661) reported A100 local compute of $13.5$~ms and a Gatherv of $22.3$~ms.
Here the same ranks, rows and kernel took $2.2$--$2.3$~ms and $2.5$--$4.0$~ms. We reran the companion script on one A100
node under four conditions, with a sampler logging each GPU's clock and the processes running on it. The kernel hash
(\texttt{e4499eb0\ldots}), torch 2.8.0, Triton 3.4.0 and mpi4py 4.1.2 were identical across the reruns; the original job's
own kernel hash was not recorded.

\begin{table}[h]
\centering
\renewcommand{\arraystretch}{1.25}
\small
\begin{tabular}{@{}p{5.6cm}p{2.4cm}rrr@{}}
\toprule
\textbf{Run} & \textbf{GPUs in use} & \textbf{A100 (ms)} & \textbf{CPU (ms)} & \textbf{Gatherv (ms)} \\
\midrule
\rowcolor{moaLightOrange} Original (22603661), \texttt{mpirun}, default & not sampled & 13.54 & 6.3 & 22.3 \\
\rowcolor{moaLightOrange} A (22764787), same script & \textbf{1} (4 PIDs) & 10.51 & 6.8 & 10.9 \\
\rowcolor{moaLightBlue} B (22764788), ob1/tcp, \texttt{vader} & 4 & 2.33 & 6.8--6.9 & 2.5--2.9 \\
\rowcolor{moaLightBlue} C (22765023), default, bound & 4 & 2.97 & 6.7 & 13.9 \\
\rowcolor{moaLightTeal} \texttt{srun}, ob1/tcp, stage 2 (22765070) & 4 & 2.33 & 6.7 & 2.5 \\
\bottomrule
\end{tabular}
\caption{The same eight-rank job under four launch conditions. A: the sampler saw one GPU UUID and four process IDs.
B excludes UCX, UCC and the CUDA shared-memory component; C leaves the default transport but sets
\texttt{CUDA\_VISIBLE\_DEVICES} per rank before Python starts. Rank~0 of B ran at 2.97~ms (probably a lower clock).
One job each, except the \texttt{srun} row, repeated twice: A100 $2.33$, $2.35$, $2.35$~ms; CPU $6.74$, $6.70$, $6.70$~ms;
Gatherv $2.50$, $2.82$, $2.98$~ms. The other rows have no between-run spread.}
\label{tab:onegpu}
\end{table}

\begin{center}
\resizebox{0.95\linewidth}{!}{%
\begin{tikzpicture}[x=1cm,y=1cm,font=\small]
\def\k{0.30}
\foreach \i/\n/\a/\g in {0/Original/13.54/22.32, 1/A: one GPU/10.51/10.93, 2/B: ob1/2.34/2.91, 3/C: bound/2.97/13.88, 4/srun ob1/2.33/2.50}{
  \node[anchor=east,font=\scriptsize] at (-0.1,{-\i*1.1}) {\n};
  \fill[moaBlue!45,draw=moaBlue] (0,{-\i*1.1+0.05}) rectangle ({\a*\k},{-\i*1.1+0.4});
  \node[anchor=west,font=\scriptsize] at ({\a*\k+0.1},{-\i*1.1+0.22}) {\a};
  \fill[moaOrange!45,draw=moaOrange] (0,{-\i*1.1-0.4}) rectangle ({\g*\k},{-\i*1.1-0.05});
  \node[anchor=west,font=\scriptsize] at ({\g*\k+0.1},{-\i*1.1-0.22}) {\g};}
\draw[->,moaGray] (0,-5.0) -- (7.6,-5.0);
\foreach \x in {0,5,10,15,20}{\draw[moaGray] ({\x*\k},-4.9) -- ({\x*\k},-5.1) node[below,font=\scriptsize]{\x};}
\node[font=\scriptsize,anchor=west] at (7.7,-5.0) {ms};
\fill[moaBlue!45,draw=moaBlue] (5.6,-2.15) rectangle (5.9,-1.85); \node[anchor=west,font=\scriptsize] at (6.0,-2.0) {A100 compute};
\fill[moaOrange!45,draw=moaOrange] (5.6,-2.65) rectangle (5.9,-2.35); \node[anchor=west,font=\scriptsize] at (6.0,-2.5) {Gatherv};
\end{tikzpicture}%
}
\end{center}
\begin{center}
\small\color{moaGray} A100 compute (blue) and Gatherv (orange) for the same eight-rank job under five launch conditions. One GPU
(A) slows the compute; the default transport (original, A, C) slows the Gatherv.
\end{center}

\textbf{Two separate effects.} (1) \emph{GPU sharing.} In A the sampler shows one GPU UUID and four PIDs: the four
A100 ranks time-sliced one device, giving about $4\times2.2$~ms plus switching ($10.5$~ms). With the GPUs separated, by
excluding the default transport (B) or by binding explicitly (C), each rank gets its own A100. A plausible mechanism,
\emph{not established here}, is that CUDA is initialized inside \texttt{MPI\_Init}, before the driver script sets
\texttt{CUDA\_VISIBLE\_DEVICES}, so every rank lands on GPU~0; we have not identified which Open~MPI component does
it. (2) \emph{Transport.} The default transport makes Gatherv slower even with four GPUs (C: $13.9$~ms against $2.5$--$2.9$~ms
under ob1/tcp). The original $22.3$~ms is higher than either, and we cannot split it between the two effects because
that job had no sampler. The CPU ranks are unaffected ($6.3$--$6.9$~ms in every row), so the CPU floor stands.

\textbf{Clocks.} Samples of the A100s show $1410$~MHz in some jobs and $1095$~MHz in others, and and the one job (C)
whose GPUs were sampled at $1095$~MHz took $2.97$~ms, $1.29\times$ the $2.3$~ms of the jobs sampled at $1410$~MHz. That
suggests compute time follows roughly $1/\mathrm{clock}$, but the evidence is weak: the samplers are one second apart and in
the repeat jobs the $1095$~MHz samples fall in the seconds after the work ended, while the compute times ($2.20$--$2.22$ and
$2.32$--$2.35$~ms) stayed the same as at $1410$~MHz. We therefore treat the clock effect as a hypothesis. Rank~0 of B
(2.97~ms) and all four ranks of C remain unexplained by anything else we measured.

\textbf{What this changes.} The A100's non-linearity is the tile-quantization staircase of Section~\ref{sec:tiles}, and
not a per-invocation overhead. The key/value-split run (Section~\ref{sec:kvsplit}) had no sampler the first time. Its two repeats did: both
show four distinct GPU UUIDs with four process IDs, so the binding there is verified (\texttt{srun}, ob1/tcp).

% ======================================================================
\section{Why Time Is Not Linear in Rows: Launch Geometry, Tested}
\label{sec:tiles}
% ======================================================================

The weight formula treats a device as a rate. A GPU kernel is not a rate: it launches a fixed number of
\emph{programs} (Triton) or \emph{teams} (OpenMP offload), one per tile of the output, and the device runs
them in waves. The number of tiles is a function of the shapes, which is exactly the kind of quantity MoA
already derives, so the fixed cost can be \emph{predicted} instead of fitted.

\begin{table}[h]
\centering
\renewcommand{\arraystretch}{1.3}
\small
\begin{tabular}{@{}p{2.3cm}p{4.1cm}p{3.3cm}p{3.3cm}@{}}
\toprule
\textbf{Device} & \textbf{Launch geometry (read from the kernel)} & \textbf{Full problem (8192 rows)} &
\textbf{This run's slice} \\
\midrule
\rowcolor{moaLightBlue} A100 (Triton) & $B{\cdot}H\cdot\lceil r/64\rceil$ programs, fixed $64\times64$ tile; 108 SMs &
256 programs & 1984 rows: 62 programs \\
\rowcolor{moaLightOrange} MI100 (OpenMP) & $B\cdot\lceil r/32\rceil$ teams of 10 workers, each walking all 512
column tiles in order & 512 teams & 6 rows: $2\cdot1=2$ teams \\
\bottomrule
\end{tabular}
\caption{Launch geometry read from the two GPU kernels. The number of waves a device needs
is $\lceil \text{programs}/\text{capacity}\rceil$, so time is a staircase in the tile count, not a line in the
row count.}
\label{tab:geometry}
\end{table}

\textbf{The A100 test.} Before the scan, the model (one program resident per SM, 108 SMs) made a sharp
prediction: the A100 time should be flat from 61 to 3456 rows ($B{\cdot}H{\cdot}54=108$ programs), step at 3520
rows, and step again at 6976 rows, where a linear rule predicts a smooth ramp. The scan then measured one A100
alone, 20 repetitions per row count (kernel hash \texttt{e4499eb0\ldots}, torch 2.8.0+cu128, Triton 3.4.0).
The time is flat at $2.18$--$2.33$~ms from 128 to 3456 rows, steps to $4.03$--$4.28$~ms from 3520 rows, and
to $6.02$--$6.09$~ms from 6976 rows: one, two and three waves, with both steps exactly where predicted. The
heights were predicted less well: the prediction used $6.874$~ms for the full problem and so expected $4.6$ and
$6.9$~ms, about 12\% above the $4.1$--$4.3$ and $6.09$~ms measured. A straight line through the same points
has $R^2=0.78$. One result was not predicted: in the first scan 61 rows took $2.88$~ms, about 25\% above the 128-row plateau. It
reproduced on two other nodes ($2.87$ and $2.90$~ms), where the 128-row point was also high ($2.76$, $2.67$~ms) and, in one repeat, the 256-row
point too ($2.68$~ms). Every elevated point is among the first few of a scan, after the GPU had idled at about $210$~MHz. A dedicated test then
confirmed a \emph{warm-up transient}. Started cold, 61, 128 and 256 rows took $2.88$, $2.67$ and $2.67$~ms and then settled; repeating the same three
sizes right afterwards gave $2.27$, $2.15$ and $2.14$~ms. After 3 seconds of the full-size problem first, the first 61, 128 and 256 rows took $2.25$,
$2.14$ and $2.14$~ms, with no slow point. A second cold start reproduced the first ($2.88$, $2.74$, then $2.15$). So the 61-row ``anomaly'' is not a property of
61 rows, and the first points of a scan carry a warm-up bias of up to about 30\%; the number of slow points varies from run to run. The clock log, now lined up with the timings, agrees with a clock ramp. The GPU sat at $210$~MHz until the process started, rose to $1095$~MHz
before the first timed point, and reached its $1410$~MHz maximum only during the 512-row point (whose first repetition was $2.71$~ms against a $2.19$~ms median). The three slow cold points therefore ran at
$1095$~MHz. They were $1.27$, $1.25$ and $1.25$ times slower than the same sizes warm, against a clock ratio of $1.29$, slightly less, as expected if part
of the kernel does not scale with the core clock. In the pre-warmed run the clock was at $1410$~MHz from before the first point and no point was slow; in the second cold start the 128-row point
(the one that overlapped the rise to $1410$~MHz) was the intermediate one. This is one device and three starts, and a sampler every
$0.22$~s, so it supports the clock ramp without isolating other first-launch effects. A small residual remains: 61 rows ($2.27$~ms warm) is still about 6\% above 128 rows ($2.14$~ms). The plateau, the two steps and the heights above $3456$ rows reproduce to within $0.4\%$, and the
CPU curve to within $1.1\%$ (from $6.2$~ms at 61 rows to $217$~ms at 8192).

\begin{table}[h]
\centering
\renewcommand{\arraystretch}{1.25}
\small
\begin{tabular}{@{}rrrrrr@{}}
\toprule
\textbf{Query rows} & \textbf{A100 alone} & \textbf{A100 beside 8} & \textbf{CPU alone} & \textbf{CPU beside 8} &
\textbf{MI100} \\
\midrule
\rowcolor{moaLightBlue} 128 & 2.29 & 2.77 & 6.22 & 6.75 & 727.9 \\
1024 & 2.20 & 2.21 & 30.9 & 36.6 & 718.2 \\
\rowcolor{moaLightBlue} 2048 & 2.21 & 2.22 & 55.9 & 66.6 & 725.7 \\
3456 & 2.18 & 2.19 & 93.1 & 108.3 & --- \\
\rowcolor{moaLightOrange} 3520 & 4.03 & 4.05 & 93.0 & 108.8 & --- \\
4096 & 4.10 & 4.08 & 111.7 & 129.2 & 1431.4 \\
6912 & 4.28 & 4.28 & 180.2 & 212.0 & --- \\
\rowcolor{moaLightOrange} 6976 & 6.02 & 6.03 & 186.4 & 215.8 & --- \\
8192 & 6.09 & 6.09 & 217.4 & 253.4 & 2128.6 \\
\bottomrule
\end{tabular}
\caption{Median time per call in milliseconds from the row scans (20 repetitions per point for the A100 and CPU,
5--14 for the MI100). ``Beside 8'' means all nine Phase~3 ranks running at once; the MI100 was scanned on its
own node. Shaded rows bracket the two predicted A100 steps. One job per device, so no between-run spread.}
\label{tab:scans}
\end{table}

\begin{center}
\resizebox{0.78\linewidth}{!}{%
\begin{tikzpicture}[x=0.00105cm,y=0.62cm,font=\small]
\draw[->,moaGray] (0,0) -- (8800,0) node[right,font=\scriptsize]{rows};
\draw[->,moaGray] (0,0) -- (0,7.6) node[above,font=\scriptsize]{ms};
\foreach \xx in {0,2048,4096,6144,8192}{\draw[moaGray] (\xx,0) -- (\xx,-0.12) node[below,font=\scriptsize]{\xx};}
\foreach \yy in {2,4,6}{\draw[moaGray] (0,\yy) -- (-90,\yy) node[left,font=\scriptsize]{\yy};}
\draw[moaGray,dashed,very thick] (0,0) -- (8192,6.088);
\node[moaGray,font=\scriptsize,anchor=west,align=left] at (6300,2.0) {linear rule\\($w_i\propto1/t_i$)};
\draw[moaBlue,very thick] (0,2.2) -- (3456,2.2) -- (3456,4.15) -- (6912,4.15) -- (6912,6.09) -- (8192,6.09);
\node[moaBlue,font=\scriptsize,anchor=south west,align=left] at (150,4.9) {tile model: step positions predicted\\before measuring (heights drawn from the data)};
\foreach \xx/\yy in {61/2.878,128/2.293,256/2.304,512/2.327,1024/2.200,1984/2.212,2048/2.210,3456/2.179,3520/4.034,4096/4.095,6912/4.279,6976/6.019,8192/6.088}{\fill[moaOrange] (\xx,\yy) circle (3pt);}
\node[font=\scriptsize,anchor=north west,align=left] at (3900,1.3) {measured, one A100 alone (orange)};
\end{tikzpicture}%
}
\end{center}
\begin{center}
\small\color{moaGray} A100 time against query rows: the straight line the weight formula assumes (grey dashed), the
staircase the launch geometry predicts (blue), and the 13 measured points (orange). The two steps fall where the
model said; the extra height at 61 rows turned out to be a warm-up transient (Section~\ref{sec:tiles}).
\end{center}

\textbf{The CPU is the other shape.} One 15-thread CPU rank is nearly linear in rows ($4.3$~ms $+\,25.9~\mu$s per
row, $R^2=0.9996$) above a floor of $6.2$~ms that holds from 61 to 128 rows. With all nine ranks running, the CPU
ranks are about 17--19\% slower (66.6 against 55.7~ms at 1984 rows, 253 against 217~ms at 8192 rows), and their
spread at 8192 rows is wide ($227$--$301$~ms); the A100 ranks are unaffected ($2.22$ against $2.21$~ms). The cause
of the CPU slowdown is not identified.

\textbf{The MI100.} The model predicted $t(6)\approx t(32)$ (a linear rule predicts a $5\times$ difference). The scan
measured $599$, $605$ and $610$~ms at 6, 32 and 64 rows; the 6-row interquartile range ($594$--$609$~ms) covers all
three, so the prediction holds, and 64 rows (4 teams) shows no sign of the extra teams. The 6-row time also
reproduces between jobs ($597.0$, $598.9$, $602.7$ and $600.2$~ms over four runs, within $1\%$). Two later repeats of the whole scan
(on \texttt{gpud01}, after the GPU request was fixed to ask for an MI100; see below) agree at every row count to within $1.5\%$: $603.7$ and
$602.1$~ms at 32 rows, $724.0$ and $723.6$~ms at 128, $726.3$ and $724.1$~ms at 2048, $1430$ and $1419$~ms at 4096, and $2129$ and
$2127$~ms at 8192. The floor, the plateau and the $3\times$ at 8192 rows are therefore properties of this device and kernel, not of one run. Beyond that the model does not explain the curve.
At 128 rows the time steps up by $118$~ms to $728$~ms, and then stays at $718$--$729$~ms through 2048 rows (from
2 to 128 teams). It doubles at 4096 rows ($1431$~ms) and reaches $2129$~ms at 8192 rows, which is three times the
plateau although rows quadrupled from 2048. A launch-geometry model that counts teams cannot produce a floor
of $600$~ms that does not move while the team count grows 64-fold, nor a $3\times$ for a $4\times$ row increase.
A floor-decomposition run (\texttt{gpud01}, verified \texttt{gfx908}, 6 rows) then tested the explanation we had written down beforehand, that the floor is
a latency chain through the key-tile loop. It ran only $T$ of the 512 key tiles (Table~\ref{tab:floor}). Time is linear in $T$,
$t\approx1.5+1.18\,T$~ms ($R^2=0.9997$, largest residual $8$~ms), so the floor is neither launch nor data transfer: with no key tile at all the call takes $2.2$~ms, and
nearly all of the $611$~ms is the key loop, at about $1.2$~ms per 16-key tile ($74~\mu$s per key). But the prediction that a larger tile would
shorten a latency chain failed. With the full 8192 keys, a key tile of 16, 32 and 64 took $611$, $592$ and $820$~ms. The time follows the number of keys processed, not the number of tiles,
and it does not depend on the number of rows (6 rows and 2048 rows differ by only $20\%$). So within a team the work on each key behaves as if it were nearly
serial, at a cost of about $10^5$ clock cycles per key for $4096$ multiply-adds.

Two follow-up runs (job 22817300, same device, 6 rows, all outputs checked against the reference) changed what the floor means. First, we split the 8192 keys over $K_s$ teams,
each returning a partial softmax triple (running maximum, sum and unnormalized output) that a second kernel merges, the same object the key/value split of
Section~\ref{sec:kvsplit} passes between ranks. With $K_s=1,4,16,64$ the time fell from $606$ to $185$, $48.7$ and $15.2$~ms (Table~\ref{tab:ksplit}),
$40\times$ at 64 chunks. We had predicted $155$, $40$ and $12$~ms from $2.2+609/K_s$; the measured values are 20--27\% higher and follow
$t\approx3.8+726/K_s$~ms for $K_s\ge4$. The slope $726$~ms is the old plateau ($718$--$733$~ms), and the $606$~ms of $K_s=1$ is the old floor, so the step from floor to plateau
at 128 rows looks like a per-team slowdown that appears once about eight or more teams run (it holds up to 128 teams); that reading fits all four points but is not isolated. Second, the kernel's own settings matter.
Raising the workers per team from 10 to 64 cut the time from $611$ to $194$~ms ($3.1\times$) and 256 workers gave $207$~ms, no better; with only 10 workers a team uses a small
part of a 64-lane wavefront. A narrower vector made it slower, not faster: 16, 32 and 64 lanes took $1834$, $1025$ and $611$~ms, so the explanation that the 64-lane vector was being
split into scalar code is not supported. The floor is therefore not a property of the MI100 alone but of this kernel's launch settings (10 workers, one team per row tile and batch element),
and the earlier statement that the MI100 should receive no rows applies to that configuration. How workers and key chunks combine, and what the tuned device does across the whole row scan, is shown below. Generated code and a profiler have still not been examined.
The straight-line fit ($600$~ms $+\,183~\mu$s per row,
$R^2=0.96$) looks acceptable and is misleading, because it hides the plateau and the steps.

\begin{table}[h]
\centering
\renewcommand{\arraystretch}{1.2}
\small
\begin{tabular}{@{}lrrrrrrr@{}}
\toprule
\textbf{Key tiles run, $T$ (tile 16)} & 0 & 1 & 4 & 16 & 64 & 256 & 512 \\
\midrule
\rowcolor{moaLightBlue} \textbf{Median time (ms)} & 2.2 & 3.4 & 7.2 & 21.6 & 77.8 & 296 & 611 \\
\midrule
\textbf{All 8192 keys, tile size} & \multicolumn{2}{r}{16} & \multicolumn{2}{r}{32} & \multicolumn{2}{r}{64} & \\
\rowcolor{moaLightOrange} \textbf{Median time (ms)} & \multicolumn{2}{r}{611} & \multicolumn{2}{r}{592} & \multicolumn{2}{r}{820} & \\
\bottomrule
\end{tabular}
\caption{MI100 floor decomposition at 6 rows (job 22777306, \texttt{gfx908}, 10 repetitions each). Time is linear in the keys processed, about $1.2$~ms per 16 keys, with a $2.2$~ms intercept; a larger tile does not help.}
\label{tab:floor}
\end{table}

\begin{table}[h]
\centering
\renewcommand{\arraystretch}{1.2}
\small
\begin{tabular}{@{}p{5.2cm}rrrr@{}}
\toprule
\textbf{Key chunks $K_s$} & 1 & 4 & 16 & 64 \\
\midrule
\rowcolor{moaLightOrange} Predicted before running (ms) & 611 & 155 & 40 & 12 \\
\rowcolor{moaLightBlue} \textbf{Measured, 6 rows (ms)} & 606 & 185 & 48.7 & 15.2 \\
Speedup over $K_s=1$ & 1 & $3.3\times$ & $12.4\times$ & $39.9\times$ \\
\midrule
\textbf{Workers per team, 10 / 64 / 256} & \multicolumn{4}{l}{611 / 194 / 207 ms} \\
\textbf{Vector lanes, 16 / 32 / 64} & \multicolumn{4}{l}{1834 / 1025 / 611 ms} \\
\bottomrule
\end{tabular}
\caption{MI100 key-split and setting sweep at 6 rows (job 22817300, \texttt{gfx908}, 10 repetitions each). Splitting the keys over teams removes most of the floor; the measured times follow $3.8+726/K_s$~ms and sit above the prediction.}
\label{tab:ksplit}
\end{table}

\begin{table}[h]
\centering
\renewcommand{\arraystretch}{1.2}
\small
\begin{tabular}{@{}p{4.6cm}rrrr@{}}
\toprule
\textbf{6 rows, ms} & $K_s=1$ & 4 & 16 & 64 \\
\midrule
\rowcolor{moaLightOrange} Predicted (64 workers) & 194 & 60 & 16 & 5 \\
\rowcolor{moaLightBlue} \textbf{Measured, 32 workers} & 198 & 50 & 14.6 & 7.2 \\
\rowcolor{moaLightBlue} \textbf{Measured, 64 workers} & 199 & 50 & 14.6 & 7.0 \\
\rowcolor{moaLightBlue} \textbf{Measured, 128 workers} & 200 & 50 & 14.5 & 7.0 \\
\bottomrule
\end{tabular}
\caption{Workers by key chunks at 6 rows (job 22817822, \texttt{gfx908}, 10 repetitions each). Workers beyond 32 change nothing; key chunks give $4\times$, $14\times$ and $28\times$ over one chunk, and below about $7$~ms a floor of roughly $4$~ms appears.}
\label{tab:tunedgrid}
\end{table}

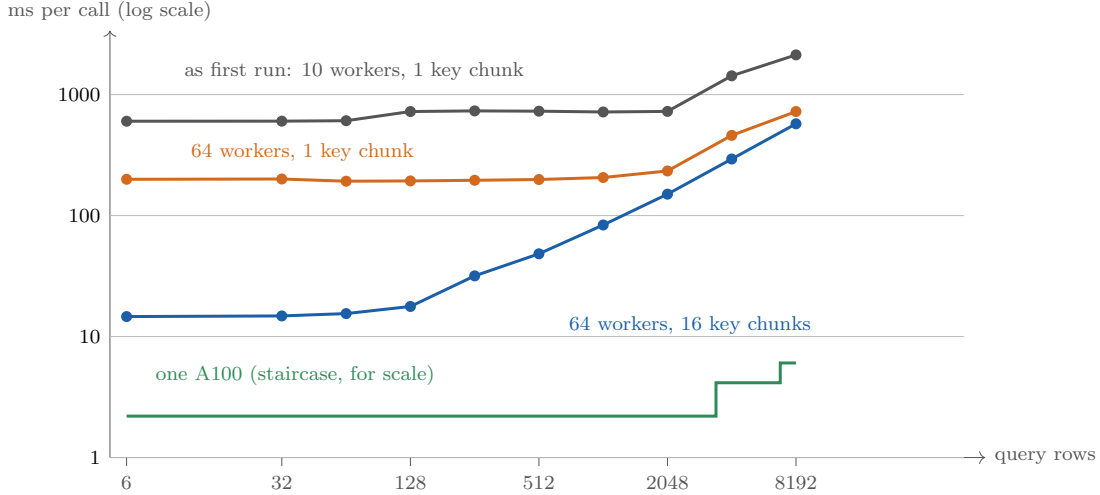
\begin{figure}[h]
\centering
\resizebox{0.86\linewidth}{!}{%
\begin{tikzpicture}[font=\small]
\draw[->,moaGray] (0,0) -- (12.3,0) node[right,font=\scriptsize]{query rows};
\draw[->,moaGray] (0,0) -- (0,6.0) node[above,font=\scriptsize]{ms per call (log scale)};
\draw[moaGray] (0.234,0) -- (0.234,-0.12) node[below,font=\scriptsize]{6};
\draw[moaGray] (2.415,0) -- (2.415,-0.12) node[below,font=\scriptsize]{32};
\draw[moaGray] (4.222,0) -- (4.222,-0.12) node[below,font=\scriptsize]{128};
\draw[moaGray] (6.028,0) -- (6.028,-0.12) node[below,font=\scriptsize]{512};
\draw[moaGray] (7.834,0) -- (7.834,-0.12) node[below,font=\scriptsize]{2048};
\draw[moaGray] (9.640,0) -- (9.640,-0.12) node[below,font=\scriptsize]{8192};
\draw[moaGray!40] (0,0.000) -- (12.0,0.000);\node[left,font=\scriptsize] at (0,0.000) {1};
\draw[moaGray!40] (0,1.700) -- (12.0,1.700);\node[left,font=\scriptsize] at (0,1.700) {10};
\draw[moaGray!40] (0,3.400) -- (12.0,3.400);\node[left,font=\scriptsize] at (0,3.400) {100};
\draw[moaGray!40] (0,5.100) -- (12.0,5.100);\node[left,font=\scriptsize] at (0,5.100) {1000};
\draw[moaGreen,very thick] (0.234,0.582) -- (8.516,0.582) -- (8.516,1.051) -- (9.419,1.051) -- (9.419,1.329) -- (9.640,1.329);
\draw[moaGray,very thick] (0.234,4.726) -- (2.415,4.727) -- (3.319,4.733) -- (4.222,4.862) -- (5.125,4.871) -- (6.028,4.867) -- (6.931,4.856) -- (7.834,4.864) -- (8.737,5.364) -- (9.640,5.658);
\fill[moaGray] (0.234,4.726) circle (2.2pt);
\fill[moaGray] (2.415,4.727) circle (2.2pt);
\fill[moaGray] (3.319,4.733) circle (2.2pt);
\fill[moaGray] (4.222,4.862) circle (2.2pt);
\fill[moaGray] (5.125,4.871) circle (2.2pt);
\fill[moaGray] (6.028,4.867) circle (2.2pt);
\fill[moaGray] (6.931,4.856) circle (2.2pt);
\fill[moaGray] (7.834,4.864) circle (2.2pt);
\fill[moaGray] (8.737,5.364) circle (2.2pt);
\fill[moaGray] (9.640,5.658) circle (2.2pt);
\draw[moaOrange,very thick] (0.234,3.910) -- (2.415,3.914) -- (3.319,3.883) -- (4.222,3.887) -- (5.125,3.896) -- (6.028,3.907) -- (6.931,3.935) -- (7.834,4.027) -- (8.737,4.527) -- (9.640,4.862);
\fill[moaOrange] (0.234,3.910) circle (2.2pt);
\fill[moaOrange] (2.415,3.914) circle (2.2pt);
\fill[moaOrange] (3.319,3.883) circle (2.2pt);
\fill[moaOrange] (4.222,3.887) circle (2.2pt);
\fill[moaOrange] (5.125,3.896) circle (2.2pt);
\fill[moaOrange] (6.028,3.907) circle (2.2pt);
\fill[moaOrange] (6.931,3.935) circle (2.2pt);
\fill[moaOrange] (7.834,4.027) circle (2.2pt);
\fill[moaOrange] (8.737,4.527) circle (2.2pt);
\fill[moaOrange] (9.640,4.862) circle (2.2pt);
\draw[moaBlue,very thick] (0.234,1.981) -- (2.415,1.990) -- (3.319,2.023) -- (4.222,2.123) -- (5.125,2.553) -- (6.028,2.863) -- (6.931,3.268) -- (7.834,3.701) -- (8.737,4.195) -- (9.640,4.690);
\fill[moaBlue] (0.234,1.981) circle (2.2pt);
\fill[moaBlue] (2.415,1.990) circle (2.2pt);
\fill[moaBlue] (3.319,2.023) circle (2.2pt);
\fill[moaBlue] (4.222,2.123) circle (2.2pt);
\fill[moaBlue] (5.125,2.553) circle (2.2pt);
\fill[moaBlue] (6.028,2.863) circle (2.2pt);
\fill[moaBlue] (6.931,3.268) circle (2.2pt);
\fill[moaBlue] (7.834,3.701) circle (2.2pt);
\fill[moaBlue] (8.737,4.195) circle (2.2pt);
\fill[moaBlue] (9.640,4.690) circle (2.2pt);
\node[moaGray,font=\scriptsize,anchor=south,align=center] at (3.434,5.172) {as first run: 10 workers, 1 key chunk};
\node[moaOrange,font=\scriptsize,anchor=south,align=center] at (2.706,4.031) {64 workers, 1 key chunk};
\node[moaBlue,font=\scriptsize,anchor=north west,align=left] at (6.319,2.124) {64 workers, 16 key chunks};
\node[moaGreen,font=\scriptsize,anchor=south west,align=left] at (0.509,0.882) {one A100 (staircase, for scale)};
\end{tikzpicture}%
}
\caption{MI100 time against query rows for three kernel configurations (job 22817822, same device as the first scan), with the A100 staircase for scale. The first-run curve (grey) has the 600~ms floor and the plateau; 64 workers per team (orange) lowers the whole curve about $3\times$ and removes the step at 128 rows; 16 key chunks per row tile (blue) removes the floor, giving $14.6$~ms at 6 rows. Above a few thousand rows all three still grow roughly in proportion to rows, and the A100 remains about $95\times$ faster at 8192 rows.}
\label{fig:mi100tuned}
\end{figure}

\textbf{Tuned configuration.} A grid of workers per team (32, 64, 128) by key chunks (1, 4, 16, 64) at 6 rows (Table~\ref{tab:tunedgrid}) shows that the two gains compose until a floor
of about $4$~ms: 32, 64 and 128 workers are indistinguishable, and the times follow roughly $4+183/K_s$~ms. We predicted $194$, $60$, $16$ and $5$~ms for $K_s=1,4,16,64$; the first
three hold or are beaten, and $K_s=64$ ($7.0$~ms) misses by $40\%$, where the merge kernel, the launch and the whole-array maps (about $2.2$~ms with no work at all) set the floor. A narrower vector is slower at 64 workers
too ($310$ and $546$~ms for 32 and 16 lanes against $199$~ms for 64), as predicted. Full scans (Figure~\ref{fig:mi100tuned}) of two tuned configurations give a different picture from the first run.
With 64 workers and one key chunk the curve is flat at $192$--$207$~ms from 6 to 1024 rows and the step at 128 rows has gone, so that step belonged to the 10-worker configuration; it reaches $234$, $460$ and $725$~ms at
2048, 4096 and 8192 rows (we had predicted a plateau near $240$~ms and $700$~ms at 8192: both hold). With 64 workers and 16 key chunks the time is $14.6$~ms at 6 rows, $17.7$~ms at 128 rows and $574$~ms at 8192 rows. We had predicted this
configuration would match the one-chunk curve within 20\% at 2048 rows and above; it does not, being $36\%$ faster at 2048 and 4096 rows ($150$ and $294$~ms against $234$ and $460$~ms) and $21\%$ faster at 8192, which is a miss, probably because 16
times as many teams keep the 120 compute units busier, but we did not isolate that. All outputs match the reference to $5\times10^{-8}$ or better.

What the tuned MI100 means for the ensemble is computed, not run. At 8192 rows the A100 takes about $6$~ms, so the MI100 is still about $95\times$ slower there, and at 6 rows it is $14.6$ against $2.2$~ms for an A100 and $6.2$~ms for a CPU rank.
On the measured curves the min-max split still gives the MI100 no rows (slowest rank $2.2$~ms), because even the best configuration cannot finish any slice inside $2.2$~ms. The $1/t$ weight rule, which gave the MI100 6 rows and a $\approx600$~ms iteration, would now give it about 170 rows and
an iteration near $22$~ms by a self-consistent version of the rule (not the exact table of Section~\ref{sec:weights}), so the penalty of using the wrong rule falls from about $270\times$ to $10\times$. The conclusion that a $1/t$ split is wrong for this device stands; the
size of the penalty, and the claim that the MI100 should be idle, depended on the kernel configuration. The ensemble has not been rerun with the tuned kernel.

\begin{center}
\resizebox{0.92\linewidth}{!}{%
\begin{tikzpicture}[x=1cm,y=1cm,font=\small]
\def\h{0.0021}
\draw[->,moaGray] (-0.3,0) -- (12.6,0) node[right,font=\scriptsize]{rows};
\foreach \i/\r/\t in {0/6/599,1/32/605,2/64/610,3/128/728,4/256/729,5/512/729,6/1024/718,7/2048/726,8/4096/1431,9/8192/2129}{
  \fill[moaOrange!45,draw=moaOrange] ({0.35+\i*1.22},0) rectangle ({0.35+\i*1.22+0.9},{\t*\h});
  \node[font=\scriptsize] at ({0.35+\i*1.22+0.45},-0.35) {\r};
  \node[font=\scriptsize] at ({0.35+\i*1.22+0.45},{\t*\h+0.25}) {\t};}
\foreach \i/\y in {0/0.0033,1/0.0175,2/0.035,3/0.07,4/0.14,5/0.28,6/0.56,7/1.12,8/2.24,9/4.47}{
  \draw[moaGray,very thick] ({0.35+\i*1.22},\y) -- ({0.35+\i*1.22+0.9},\y);}
\node[font=\scriptsize,anchor=west,color=moaGray] at (0.3,3.7) {grey tick: linear rule (weight formula)};
\node[font=\scriptsize,anchor=west,color=moaOrange] at (0.3,3.35) {bars: measured ms per call};
\end{tikzpicture}%
}
\end{center}
\begin{center}
\small\color{moaGray} MI100 time per call against query rows (bars, ms) and the linear rule from the full-problem
time (grey ticks, almost invisible below 1024 rows). A floor near 600~ms, a plateau near 725~ms, then
$2\times$ and $3\times$ rather than $4\times$.
\end{center}

\textbf{What the measured curves do to the split.} The weight formula $w_i\propto1/t_i$ rests on each $t_i$ being a
rate. Replacing it with the measured curves, and choosing the row counts that minimize the slowest rank
(a min-max split), gives a very different answer. Because the A100 is flat to 3456 rows, any share up to 3456
rows per A100 costs the same.

\begin{table}[h]
\centering
\renewcommand{\arraystretch}{1.25}
\small
\begin{tabular}{@{}p{5.3cm}p{4.6cm}p{3.1cm}@{}}
\toprule
\textbf{Split} & \textbf{Rows} & \textbf{Slowest rank} \\
\midrule
\rowcolor{moaLightOrange} Phase~3 as run ($w_i\propto1/t_i$, nine ranks) & A100 1984, CPU 61, MI100 6 &
$\approx601$~ms with Gatherv (measured) \\
\rowcolor{moaLightBlue} $w_i\propto1/t_i$ on the measured A100 and CPU curves & A100 1999, CPU 48 &
6.60~ms (computed) \\
\rowcolor{moaLightBlue} Same family, run as an ensemble (A100 1986--1990, CPU 61; \texttt{srun}, ob1) & 8 ranks &
6.74~ms compute, $9.24$~ms with Gatherv (measured) \\
\rowcolor{moaLightTeal} Min-max on the same curves & A100 2048 each, CPU 0 & 2.19~ms (computed) \\
\rowcolor{moaLightTeal} Min-max, run as an ensemble & 4 A100 ranks & 2.21~ms compute, $4.66$~ms with Gatherv (measured) \\
\bottomrule
\end{tabular}
\caption{Splits of the 8192 rows. Rows marked computed come from separately measured curves; rows marked measured are
single ensemble jobs (22765070), first-run values (two repeats are quoted in the text); the 6.74 and 9.24~ms row uses the same 8-rank split as
the companion paper, on the same kind of node. The CPU and MI100 ranks receive no rows because their smallest measured time ($6.2$~ms
and about $600$~ms) already exceeds what the A100s need.}
\label{tab:splits}
\end{table}

The honest reading is that, at this problem size, the CPU and MI100 ranks contribute no speed: three times faster than
the $1/t$ split is available by leaving them out, and the A100-only ensemble confirmed it: slowest compute
$6.74\to2.21$~ms ($3.05\times$ in the first run; $6.70\to2.22$ and $6.70\to2.22$~ms in two repeats), compute plus Gatherv
$9.24\to4.66$, $9.50\to5.00$ and $9.68\to9.14$~ms: about $1.9\times$ twice, and $1.06\times$ once, because one A100-only Gatherv took $6.9$~ms against $2.4$--$2.8$~ms in the others. They begin to matter only when the other ranks' time grows
to the order of their floors, which is a statement about problem size that the weight formula cannot make and
the staircase can. This is why the time-to-rows relation has to be derived from shapes: the shapes fix the tile
count, and the tile count fixes where the steps fall.

% ======================================================================
\section{Measuring $t_{\mathrm{comm}}$: the Communication Sweep}
\label{sec:comm}
% ======================================================================

One het job (A100 node \texttt{gpua034}, MI100 node \texttt{gpud01}, the same nine ranks; the table below), repeated twice (the second repeat on A100 node \texttt{gpua051}; see the repeat paragraph), timed point-to-point
ping-pong within a node and across the two partitions, Allreduce, and Gather over message sizes from 8~B to
64~MiB, three times: with Open MPI free to choose among the interfaces both nodes share (\texttt{auto}) and
pinned to each (\texttt{hsn0}, \texttt{hsn0.561}). Transport was \texttt{ob1} over TCP, UCX excluded. Both nodes
also list a \texttt{cxi0} device, which as far as we know is the Slingshot NIC; the TCP numbers below therefore
use only a small part of what the fabric can do.

\begin{table}[h]
\centering
\renewcommand{\arraystretch}{1.25}
\small
\begin{tabular}{@{}p{4.9cm}rrr@{}}
\toprule
\textbf{Operation, size} & \textbf{auto} & \textbf{hsn0} & \textbf{hsn0.561} \\
\midrule
\rowcolor{moaLightBlue} Ping-pong across nodes, 8 B (one way) & 0.032 & 0.025 & 0.023 \\
Ping-pong across nodes, 8 MiB & 3.49 & 4.16 & 4.62 \\
\rowcolor{moaLightBlue} Ping-pong across nodes, 64 MiB & 24.4 & 33.8 & 34.4 \\
Ping-pong within a node, 8 MiB & 1.54 & 1.59 & 1.76 \\
\rowcolor{moaLightOrange} Allreduce, 1 MiB, 9 ranks & 2.11 & 2.22 & 2.00 \\
\rowcolor{moaLightOrange} Allreduce, 8 MiB, 9 ranks & 11.1 & 11.4 & 12.4 \\
\rowcolor{moaLightTeal} Gather, 8 MiB total, 9 ranks & 1.44 & 1.18 & 1.36 \\
\bottomrule
\end{tabular}
\caption{Median time in milliseconds, first job only. Large-message bandwidth across nodes is $2.75$~GB/s (auto) and
$1.95$--$1.99$~GB/s (pinned); within a node about $6$--$6.7$~GB/s.}
\label{tab:comm}
\end{table}

\textbf{Between runs.} Two repeats changed the picture. The 8~MiB cross-node ping-pong took $3.5$, $3.3$, $8.0$~ms (auto),
$4.2$, $6.8$, $6.6$~ms (hsn0) and $4.6$, $6.4$, $5.0$~ms (hsn0.561) over the three jobs; at 64~MiB it took $24$, $53$, $31$~ms (auto),
$34$, $103$, $44$~ms (hsn0) and $34$, $68$, $36$~ms (hsn0.561), a factor of up to $4.3$ between runs of the same
configuration. The $8$~MiB Allreduce took $11.1$--$15.4$~ms (auto), $11.4$--$22.9$~ms (hsn0) and $12.0$--$12.4$~ms
(hsn0.561). Within a node and for the Gather the times are steady ($1.3$--$1.7$~ms for an $8$~MiB ping-pong; $1.2$--$1.6$~ms for the Gather).
The first job's finding that letting Open MPI choose gives about 40\% more cross-node bandwidth than either interface
alone is therefore \textbf{withdrawn}: the differences between interfaces are smaller than the differences between runs.
The cross-node numbers also depend on which other jobs share the fabric, which we did not control or measure.
For the first job's numbers, the Allreduce reaches only $0.56$--$0.72$~GB/s effective
bandwidth (message bytes over time), about $3.8\times$ below the ping-pong figure; the ring factor
$2(P{-}1)/P=1.78$ explains less than half of that, and the rest is unexplained. The linear fits
$t=\alpha+\mathrm{bytes}/\beta$ are rough: latency is not linear in bytes (64~KiB moves at $0.56$~GB/s, 1~MiB
at $2.4$~GB/s), and two pinned-interface fits returned negative $\alpha$, so small-message times should be
taken from the measured points, not from the fit. The sweep's 8.4~MB Gather took $1.2$--$1.4$~ms, against the
$4.03$~ms Gatherv of Section~\ref{sec:phase3}; the two are not directly comparable (equal blocks and a
different root and payload here, uneven blocks there). The sweep and Section~\ref{sec:phase3} both used \texttt{ob1}
over TCP. Under Open~MPI's default transport the same Gatherv took $10.9$--$22.3$~ms (Section~\ref{sec:onegpu}), so any
$t_{\mathrm{comm}}$ must be quoted together with the transport. There is one job, so
no between-run spread, and the difference between \texttt{auto} and the pinned settings could still be a
run-to-run effect.

% ======================================================================
\section{Splitting the Contraction Index: Communication Derived from Shape}
\label{sec:kvsplit}
% ======================================================================

So far MPI has carried only the final gather of results; the attention itself needed no communication, because
rows were split and every softmax stayed inside one rank. The theory has not yet been tested on a computation
whose own DNF forces communication. Splitting the \emph{key/value} axis does, and MoA lets us derive what
must be communicated from the shapes alone.

\textbf{The derivation.} $Q,K,V$ have shape $(B{\cdot}H,\,N,\,D)$. Cut the key/value axis into slices, one per rank.
Each rank then holds only part of every softmax: for each of the $N$ query rows it computes a partial triple
\[
m_r=\max_{j\in\text{slice }r} s_j,\qquad l_r=\sum_{j\in r} e^{s_j-m_r},\qquad O_r=\sum_{j\in r} e^{s_j-m_r}\,v_j,
\]
with shapes $(B{\cdot}H,N)$, $(B{\cdot}H,N)$, $(B{\cdot}H,N,D)$. These combine by the \emph{same} rescaling step the
online-softmax ONF of Paper~1 already uses between tiles inside one kernel, now applied between ranks:
\[
M=\max_r m_r,\qquad L=\sum_r l_r\,e^{m_r-M},\qquad O=\sum_r O_r\,e^{m_r-M},\qquad \text{out}=O/L .
\]
This is dimension lifting at work: the algebra that merges tiles on one device merges devices in one
communicator. Computationally it is one Allreduce(MAX) of $m$ and one Allreduce(SUM) of the packed $(l,O)$.

\begin{briefbox}[title={Worked example: two ranks, four keys}]
Scores for one query are $(1,2)$ on rank~0 and $(3,0)$ on rank~1. Rank~0 holds $m_0=2$ and
$l_0=e^{-1}+1=1.3679$; rank~1 holds $m_1=3$ and $l_1=1+e^{-3}=1.0498$. Then $M=3$ and
$L=1.3679\,e^{2-3}+1.0498\,e^{0}=0.5032+1.0498=1.5530$. Summing the four scores' exponentials against the
common maximum directly gives $e^{-2}+e^{-1}+1+e^{-3}=1.5530$: the same number, with only $(m,l)$ and a
rescale crossing the rank boundary, never the scores themselves.
\end{briefbox}

\begin{center}
\resizebox{0.97\linewidth}{!}{%
\begin{tikzpicture}[x=1cm,y=1cm,font=\small]
\def\sc{0.0013428}
\foreach \a/\b/\c in {0/1990/moaBlue!35, 1990/3976/moaBlue!22, 3976/5962/moaBlue!35, 5962/7948/moaBlue!22,
                       7948/8009/moaTeal!55, 8009/8070/moaTeal!35, 8070/8131/moaTeal!55, 8131/8192/moaTeal!35}{
  \fill[\c,draw=moaGray] ({\a*\sc},0) rectangle ({\b*\sc},0.8);}
\node[font=\small] at (5.5,1.3) {key/value axis, $N{=}8192$ keys, split $w_i\propto1/t_i$ (to scale)};
\draw[moaOrange,dashed,thick] (11.5,0) rectangle (14.3,0.8) node[midway,font=\small]{rank 8: 0 keys};
\foreach \xx/\t in {1.34/0, 4.01/1, 6.67/2, 9.34/3}{
  \node[font=\small] at (\xx,0.4) {rank \t};
  \node[font=\small,align=center] (p\t) at (\xx,-0.9) {$(m,l,O)_{\t}$};
  \draw[->,moaGray] (\xx,0) -- (p\t.north);}
\node[font=\small,align=center] (pc) at (11.3,-0.9) {$(m,l,O)_{4..7}$};
\node[font=\scriptsize,color=moaGray] at (11.3,-1.45) {61 keys each};
\draw[->,moaGray] (10.84,0) -- (pc.north);
\node[font=\small,align=center] (pm) at (13.4,-0.9) {$(-\infty,0,0)$};
\draw[->,moaGray,dashed] (12.9,0) -- (pm.north);
\node[draw=moaPurple,fill=moaLightPurple,rounded corners,align=center,minimum width=14.2cm,font=\small] (cmb) at (7.15,-3.5)
  {Allreduce(MAX) of $m$: $2\cdot8192\cdot8=131{,}072$ B \;$\Rightarrow$\; rescale by $e^{m_r-M}$ \;$\Rightarrow$\;\\[2pt]
   Allreduce(SUM) of $(l,O)$: $2\cdot8192\cdot65\cdot8=8{,}519{,}680$ B \;$\Rightarrow$\; out $=O/L$};
\foreach \xx in {1.34,4.01,6.67,9.34,11.3}{\draw[->,moaGray] (\xx,-1.7) -- (\xx,-2.95);}
\draw[->,moaGray,dashed] (13.4,-1.2) -- (13.4,-2.95);
\end{tikzpicture}%
}
\end{center}
\begin{center}
\small\color{moaGray} The key/value-split ensemble. Blue: the four A100 ranks (about 24\% of the keys each);
teal: the four CPU ranks (61 keys each); orange, dashed: the MI100 rank, which holds no keys (it has no
partial-attention kernel yet) and only joins the two reductions, so it measures the cross-partition
reduction cost and not MI100 compute.
\end{center}

\textbf{Predictions recorded before measuring.} (i) The two message sizes above are fixed by the shapes and do not
depend on how the keys are split. (ii) Output matches the fp64 reference to fp32 round-off ($\approx10^{-6}$).
(iii) A ring allreduce moves $2(P{-}1)/P\approx1.78$ times the message through each rank for $P{=}9$, so the
reduction of the $8.5$~MB packed message should cost at least $1.8\times$ what the row-split Gatherv of
the same $8.4$~MB output cost at equal bandwidth. (iv) With $t_{\mathrm{comm}}\approx\alpha+\mathrm{bytes}/\beta$ and
$(\alpha,\beta)$ taken from the separate communication sweep (per network interface, UCX excluded), the measured
reduction times should land on that line; the driver prints both. Only the Allreduce algorithm ($G_d$: recursive
doubling versus ring) is varied, set explicitly so it is a directive and not a runtime heuristic.
From the sweep we then predicted (v) about $11$--$12$~ms for the SUM and $0.3$--$0.9$~ms for the MAX with the
default algorithm.

\textbf{Results.} The job ran once for each algorithm on the same nine ranks (key counts of the figure above).
Both outputs match the fp64 reference: $\max|\mathrm{err}|=1.39\times10^{-7}$, better than the $\approx10^{-6}$
predicted in (ii). Relative errors reach about 12\% in some blocks of both outputs. The checker divides by $|\mathrm{ref}|+10^{-12}$, so a
large relative error is expected where the reference is near zero (the companion paper's check shows 6.5\% for the same
reason); we have not located the entries, so this is the likely explanation, not a verified one. The absolute errors
are the figures to use.

\begin{table}[h]
\centering
\renewcommand{\arraystretch}{1.25}
\small
\begin{tabular}{@{}p{5.0cm}rrr@{}}
\toprule
\textbf{Milliseconds (slowest rank), range of 3 jobs} & \textbf{Alg.~3 (recursive doubling)} & \textbf{Alg.~4 (ring)} &
\textbf{Predicted} \\
\midrule
\rowcolor{moaLightBlue} Slowest local partial (untuned code) & 20.5--21.5 & 19.8--21.7 & --- \\
\rowcolor{moaLightOrange} Allreduce(MAX), 131,072 B & 0.98--1.14 & 1.41--1.58 & 0.3--0.9 \\
\rowcolor{moaLightPurple} Allreduce(SUM), 8,519,680 B & 20.85--27.48 & 11.31--15.98 & 11--12 \\
Iteration (partial + both reductions) & 43.2--50.1 & 32.5--39.1 & --- \\
Effective bandwidth, SUM (GB/s) & 0.31--0.41 & 0.53--0.75 & --- \\
\bottomrule
\end{tabular}
\caption{Key/value-split ensemble: one job per algorithm in each of three submissions (the first, and two repeats on other
nodes whose GPU sampler shows four separate A100s); ranges are over the three. The figure below uses the first. The local partials are untuned: torch matrix products
on the A100 ranks (including the copy back to the host), NumPy on the CPU ranks; the MI100 holds no keys.
They are not comparable with the row-split kernels of Table~\ref{tab:phase3}.}
\label{tab:kvresults}
\end{table}

\begin{center}
\begin{tikzpicture}[x=0.2cm,y=1cm,font=\small]
\fill[moaBlue!35,draw=moaBlue] (0,0) rectangle (21.3,0.6);
\fill[moaOrange!55,draw=moaOrange] (21.3,0) rectangle (22.28,0.6);
\fill[moaPurple!40,draw=moaPurple] (22.28,0) rectangle (43.13,0.6);
\node[anchor=east,font=\scriptsize] at (-0.4,0.3) {alg.~3};
\fill[moaBlue!35,draw=moaBlue] (0,-1) rectangle (19.76,-0.4);
\fill[moaOrange!55,draw=moaOrange] (19.76,-1) rectangle (21.17,-0.4);
\fill[moaPurple!40,draw=moaPurple] (21.17,-1) rectangle (32.48,-0.4);
\node[anchor=east,font=\scriptsize] at (-0.4,-0.7) {alg.~4};
\node[anchor=west,font=\scriptsize] at (43.5,0.3) {43.2 ms};
\node[anchor=west,font=\scriptsize] at (33.0,-0.7) {32.5 ms};
\draw[->,moaGray] (0,-1.5) -- (46,-1.5) node[right,font=\scriptsize]{ms};
\foreach \x in {0,10,20,30,40}{\draw[moaGray] (\x,-1.4) -- (\x,-1.6) node[below,font=\scriptsize]{\x};}
\fill[moaBlue!35,draw=moaBlue] (3,-2.7) rectangle (4.5,-2.3); \node[anchor=west,font=\scriptsize] at (4.8,-2.5) {slowest local partial};
\fill[moaOrange!55,draw=moaOrange] (24,-2.7) rectangle (25.5,-2.3); \node[anchor=west,font=\scriptsize] at (25.8,-2.5) {Allreduce MAX};
\fill[moaPurple!40,draw=moaPurple] (38,-2.7) rectangle (39.5,-2.3); \node[anchor=west,font=\scriptsize] at (39.8,-2.5) {Allreduce SUM};
\end{tikzpicture}
\end{center}
\begin{center}
\small\color{moaGray} Where one iteration of the key/value split spends its time, for each Allreduce algorithm.
First job. The reductions are about 12.7~ms of the ring's 32.5~ms; recursive doubling nearly doubles the SUM. In the repeats
the ring's reductions took $17.6$ and $16.8$~ms of $39.1$ and $38.2$~ms.
\end{center}

\textbf{Against the predictions.} (i) Message sizes: by construction the buffers are $131{,}072$ and
$8{,}519{,}680$~B, as predicted. (ii) Correctness: passed, better than predicted. (iii) The ring reduction of
the packed message took $11.3$--$16.0$~ms against $4.03$~ms for the row-split Gatherv ($2.8$--$4.0\times$) and $1.2$--$1.6$~ms for the
sweep's Gather ($7$--$13\times$), so the $\geq1.8\times$ bound holds, loosely. (iv)/(v) The SUM prediction held once and
missed twice: the ring took $11.31$~ms in the first job (inside $11$--$12$~ms) and $15.98$ and $15.26$~ms in the repeats, 30--45\% above
the range. The sweep's own 8~MiB Allreduce also varied between runs ($11$--$15$~ms with the default interface choice), so
the prediction was only ever as good as one run of the sweep. The MAX prediction misses in every run
($0.98$--$1.14$~ms and $1.41$--$1.58$~ms against $0.3$--$0.9$~ms), plausibly because the sweep's small-message points were noisy. The algorithm matters: recursive doubling took $1.85\times$ as
long as the ring for the 8.5~MB message. A back-of-envelope count (full-buffer exchanges against $1.78$ buffer
volumes) suggests about $2\times$, but this is a rough check, not a model.

\textbf{What it means.} At $N{=}8192$ the packed partial $(l,O)$ is as large as the output itself, and the
reduction moves $1.78$ times that through every rank, where the row split gathers the output once. With about
$13$--$18$~ms of communication against an A100 partial of $3$--$5$~ms, the key/value split is communication-bound here,
and the shapes could have told us so before running. Splitting the contraction index pays when the keys do not
fit on one device or the compute per key is much larger, not at this size. The measurement also gives $t_{\mathrm{comm}}$
its first prediction that matched once (the SUM, in one of three runs) and its first miss that reproduced (the MAX). What it
teaches is that $t_{\mathrm{comm}}$ here is a distribution, not a number: a prediction from shapes and a measured $\alpha$, $\beta$
needs a spread, and we have only three samples.

% ======================================================================
\section{Variance, Energy, and the $\varepsilon$ Family}
\label{sec:epsilon}
% ======================================================================

The closing-the-gap paper \cite{closinggap} planned a family of $\varepsilon$ variables to absorb variance from
compiler- and runtime-chosen optimizations, to be introduced only after every deterministic directive is
isolated. This run suggests the discipline for doing it: \emph{separate structure from variance}. The ratios
of Table~\ref{tab:phase3} ($1.3\times$, $4\times$, $382\times$) are not $\varepsilon$; if Section~\ref{sec:tiles} is right
they are deterministic consequences of launch geometry and should be predicted from shapes. Only what is left
after the geometric model, plus the spread across repeated runs, is $\varepsilon$:
\[
\varepsilon_d^{(c)}=\frac{t_{\mathrm{meas}}-t_{\mathrm{model}}}{t_{\mathrm{model}}},
\]
one per realization $d$ and mechanism class $c$ (vector and unroll choices left to the compiler, dispatch and
scheduling, library algorithm selection, clock and power state). We can now write a first $\varepsilon$ for the A100, where the shapes give a model. Take $t_{\mathrm{model}}=k\times2.19$~ms for $k$
waves, with $k=\lceil\text{programs}/108\rceil$ from Section~\ref{sec:shapes} (the one-wave time is the unit, so one wave has
$\varepsilon=0$ by construction and says nothing). The measured two-wave times of $4.03$--$4.27$~ms give
$\varepsilon_{\mathrm{A100}}^{\text{wave}}=-8.0\%$ to $-2.5\%$ against the predicted $4.38$~ms, and the three-wave times of $6.02$--$6.09$~ms give
$-8.4\%$ to $-7.3\%$ against $6.57$~ms. The sign is consistent (a later wave costs slightly less than the first), and the size is of the order of
the run-to-run spread plus the unexplained 12\% in the step heights; with one device, one kernel and two step counts this is a first estimate, not a
distribution. We have no $\varepsilon$ for the MI100, because its floor has been located (the key loop) but not explained, and the CPU line (a fixed $4.25$~ms plus $25.9~\mu$s per row,
$R^2=0.9996$) is a fit, not a model from shapes. Variance inputs: within a run, the A100's 20 repetitions span under 3\% at nearly every
row count (one outlier at 128 rows: $2.29$ median, $2.76$ maximum), the MI100's 6-row calls span $587$--$653$~ms
(interquartile $594$--$609$~ms), and the concurrent CPU ranks span $227$--$301$~ms at 8192 rows. Between runs (three submissions each, different nodes),
the A100 scan agrees to $0.4\%$ at every row count above 256, the CPU scan to $1.1\%$, and the MI100 scan to $1.5\%$ (6-row time $597.0$, $598.9$, $602.7$, $600.2$~ms);
the large between-run differences are all in communication (Section~\ref{sec:comm}), and the $13.5$~ms of \cite{mpiext} was a launch artifact
(Section~\ref{sec:onegpu}). So for computation $\varepsilon$ is small and stable; for communication it is not yet bounded.

\textbf{MI100 energy.} The device's own energy counter returned a zero delta under confirmed load in every run
so far \cite{closinggap}, so a $P_d$-based energy prediction for the MI100 has no measured ground truth. The scan
job therefore samples package power through \texttt{rocm-smi} (about one sample per second) around timed windows
of at least eight seconds, and estimates energy per call as the mean sampled power times the time per call. This
is an estimate and is labeled as one. From 141 power samples ($39$--$177$~W) it gives about $58$~J per call at
6 rows, $109$~J at 2048 rows and $308$~J at 8192 rows, so energy per row falls from about $9.7$~J at 6 rows to
$0.038$~J at 8192 rows as the fixed cost is amortized. The two scan repeats ran the same power sampler. Their estimates agree with the first run for 6 to 2048 rows
($56.7$ and $57.5$~J at 6 rows; $107.0$ and $106.4$~J at 2048 rows, within 3\% of the first run's $58$ and $109$~J), but not at the largest size:
$324.7$ and $294.3$~J at 8192 rows against $308$~J first, a spread of about 10\% across the three runs, and $206$ and $194$~J at 4096 rows.
Each point rests on 10--15 one-second samples, so the sampling is coarse; the large-size spread is the sampler's, and we cannot say how much of it is the device's.
The estimate is stable where power is nearly constant and loose where it is not. A later job (22817822) with the tuned kernels gives, with the same sampler, $20.0$~J per call at 6 rows with 64 workers and one key chunk
($57$~J before) and $1.8$~J with 16 key chunks, a $32\times$ drop; at 8192 rows $126$ and $101$~J against about $309$~J. At 6 rows the 16-chunk figure rests on 7 power samples. The drop comes mostly from the shorter time, since the
average power is similar ($100$--$123$~W at 6 rows) or higher. The number gives the
power model a figure to be wrong about; it is not a hardware measurement.

% ======================================================================
\section{Extending $C_d$'s Admissible Class}
\label{sec:cd-extend}
% ======================================================================

\cite{closinggap}'s admissibility criterion --- select a real hardware quantity with no discretionary
reordering, be deterministic rather than runtime-overridable --- was stated generally enough that this
paper needs no new criterion, only a new instance of it. Table~\ref{tab:admissible-extend} summarizes what
each paper in this series has added to the admissible class.

\begin{table}[h]
\centering
\renewcommand{\arraystretch}{1.3}
\small
\begin{tabular}{@{}p{3.1cm}p{5.6cm}p{4.3cm}@{}}
\toprule
\textbf{Paper} & \textbf{Admissible mechanism added} & \textbf{Layer} \\
\midrule
\rowcolor{moaLightOrange} Closing-the-gap \cite{closinggap} & cache placement, deterministic vector/tile
width, occupancy caps & compiler/runtime directive \\
\rowcolor{moaLightBlue} MPI-extension \cite{mpiext} & explicit \texttt{coll\_tuned\_gather\_algorithm}
($G_d$), NUMA-aligned rankfile placement ($L_d$) & MPI runtime directive \\
\rowcolor{moaLightPurple} This paper & Slurm het-job component specification (partition, node count, GPU
type per component) & job-orchestration directive \\
\bottomrule
\end{tabular}
\caption{The admissible class grows by layer, not by relaxing the criterion itself.}
\label{tab:admissible-extend}
\end{table}

% ======================================================================
\section{Next Steps}
% ======================================================================

\begin{enumerate}
\item \textbf{Done:} the standalone matching-shape MI100 timing test
(Section~\ref{sec:weights}): $t_{\mathrm{MI100}}=2134.261$~ms, correctness passed, device use confirmed.
\item \textbf{Done:} one nine-rank communicator (4~A100, 4~CPU, 1~MI100) across both het-job components,
its measured per-rank times, Gatherv cost ($4.03$~ms), and the independent fp64-reference check
(Section~\ref{sec:phase3}). Repeated in the A100-only and A100+CPU jobs, the kv-split, the A100/CPU scan and the comm sweep (two repeats each); compute times
reproduce to about 1\%, communication times do not. The MI100 scan was repeated twice and agrees to within $1.5\%$.
\item \textbf{Done:} the row scans on all three device types (Section~\ref{sec:tiles}). The A100 steps fall at 3520
and 6976 rows as predicted, and its warm-up transient agrees with the clock log. The MI100's $\approx600$~ms floor, plateau and $3\times$ at 8192 rows belong to its first kernel
configuration (10 workers, one team per row tile); a tuned configuration lowers the floor to $14.6$~ms. The CPU is nearly linear above a $6.2$~ms floor.
\item \textbf{Done:} the communication sweep (Section~\ref{sec:comm}) and the key/value-split ensemble
(Section~\ref{sec:kvsplit}): correct, SUM within the predicted range in one of three runs, MAX missed in all, communication-bound at this size (GPU binding verified in the repeats).
\item \textbf{Done, mechanism open:} the A100 $8.1\times$ in \cite{mpiext}'s final run ($13.5$~ms) against $1.3\times$
here. Reruns show four ranks on one GPU under the default transport (Section~\ref{sec:onegpu}). Still open: which
Open~MPI component (UCX, smcuda, UCC, hcoll) initializes CUDA before the binding, and the original job's own GPU use
(it had no sampler).
\item \textbf{Done:} the computed min-max split was run as an ensemble (A100 ranks only) and agrees with the computed
value (Table~\ref{tab:splits}).
\item \textbf{Done, with a lesson:} the first two MI100 scan repeats refused to run, because node \texttt{gpud01} holds eight MI100s
and one MI210 (\texttt{gfx90a}) and our untyped GPU request was given the MI210; the script's architecture
guard stopped them. Requesting \texttt{--gres=gpu:mi100:1} fixed it, and the repeats passed the guard and reproduced the first scan
(Section~\ref{sec:tiles}). The standalone and nine-rank runs used the untyped request and did not log the architecture; their MI100
times (2134 and 597~ms) match the verified scans to within $0.3\%$, which is strong circumstantial evidence, not proof, that they ran on an MI100.
\item \textbf{Done:} the MI100 floor decomposition, a key-split (partial-attention) kernel, the workers-by-chunks grid, two tuned row scans and the
MI100 energy repeats (Sections~\ref{sec:tiles} and~\ref{sec:epsilon}). The floor is the key loop under the first launch settings; the tuned kernel is still $95\times$ slower than an A100 at 8192 rows.
\item \textbf{Carried to the next paper (Stampede3 and the Intel Max 1550):} rerun the Phase~3 ensemble and the weight table with the tuned MI100 kernel; use the key-split
kernel as the MI100 rank in the key/value-split ensemble; find out why the first configuration's $726$~ms plateau appeared and profile the kernel; put a spread on the comm sweep with more than three runs
and a record of fabric load; and add the Intel Max 1550 as a fourth vendor.
\item Do not compare $t_{\mathrm{comm}}$ with \cite{mpiext}'s $\approx$22~ms: it was measured under the default
transport with four ranks on one GPU. Re-measure under both transports with a sampler running; ask NCSA why UCX triggers the job cancellations and whether the \texttt{cxi0}
fabric can be used (TCP reaches $2$--$2.8$~GB/s).
\item Formalize the candidate $\rho_{\mathrm{machine}}(d)$ fields the runs forced into view: the launch-geometry
cost (per-device floors of $\approx2.2$~ms per A100 wave, $6.2$~ms for the CPU and $14.6$--$600$~ms for the MI100, depending on its kernel configuration)
and the inter-node transport ($\alpha$, $\beta$ per interface and operation, Table~\ref{tab:comm}); estimate the first
$\varepsilon$ values once the model predicts heights as well as positions.
\item Tune the partials of the other ranks so the comparison between the key/value split and the row split is fair, and scale the MI100 rank count beyond
$k{=}1$; keep Intel's Max 1550 in the no-communication model until it is run under its own allocation.
\end{enumerate}

% ======================================================================
\section*{Acknowledgments}
% ======================================================================

This work used NCSA Delta (GPU and CPU partitions), TACC Stampede3, PSC Bridges-2 (GPU and Ocean Storage),
SDSC Expanse (GPU and CPU partitions), and Purdue Anvil (GPU) through allocation(s) CIS261342 and
CIS261396 from the Advanced Cyberinfrastructure Coordination Ecosystem: Services \& Support (ACCESS)
program, which is supported by U.S. National Science Foundation grants \#2138259, \#2138286, \#2138307,
\#2137603, and \#2138296.

% ======================================================================


\begin{thebibliography}{9}

\bibitem{hetero} L.~M. Mullin and G.~Hains. One Derivation, Any Machine: A No-Communication Weighted
Partition Scheme for Heterogeneous Multi-Vendor Ensembles. Under review, target venue MLSys. 2026.

\bibitem{closinggap} L.~M. Mullin and G.~Hains. Closing the Prediction Gap: Admissible Compiler and
Runtime Directives for $\rho_{\mathrm{machine}}(d)$. 2026.

\bibitem{mpiext} L.~M. Mullin and G.~Hains. From No-Communication to MPI: Extending the
$\rho_{\mathrm{machine}}(d)$ Predictive Theory with a Real CPU+GPU Ensemble Experiment on Attention. 2026.

\end{thebibliography}
\end{document}